\documentclass[aps,preprint,superscriptaddress]{revtex4-1}

\usepackage{graphicx}
\usepackage{amsmath}
\usepackage{amssymb}
\usepackage{slashed}
\usepackage{braket}

\newcommand{\tr}{{\rm tr \,}}

\usepackage{color}

\begin{document}

\title{On the axial-vector form factor of the nucleon \\ and chiral symmetry}

\author{Matthias F.M. Lutz}
\affiliation{GSI Helmholtzzentrum f\"ur Schwerionenforschung GmbH, \\Planckstra\ss e 1, 64291 Darmstadt, Germany}
\affiliation{Technische Universit\"at Darmstadt, D-64289 Darmstadt, Germany}
\author{Ulrich Sauerwein}
\affiliation{GSI Helmholtzzentrum f\"ur Schwerionenforschung GmbH, \\Planckstra\ss e 1, 64291 Darmstadt, Germany}
\affiliation{Technische Universit\"at Darmstadt, D-64289 Darmstadt, Germany}
\affiliation{Van Swinderen Institute for Particle Physics and Gravity,University of Groningen, 9747 AG Groningen, The Netherlands}

\author{Rob G.E. Timmermans}
\affiliation{Van Swinderen Institute for Particle Physics and Gravity,University of Groningen, 9747 AG Groningen, The Netherlands}

\date{\today}

\begin{abstract}
We consider the chiral Lagrangian with nucleon, isobar, and pion degrees of freedom. The baryon masses and the axial-vector form factor of the nucleon are derived at the one-loop level. We explore the impact of using on-shell baryon masses in the loop expressions. As compared to results from conventional chiral perturbation theory we find significant differences. An application to QCD lattice data is presented.  We perform a global fit to the available lattice data sets for the baryon masses and the nucleon axial-vector form factor, and determine the low-energy constants relevant at N$^3$LO for the baryon masses and at N$^2$LO for the form factor. Partial finite-volume effects are considered. We point out that the use of on-shell masses in the loops results in non-analytic behavior of the baryon masses and the form factor as function of the pion mass, which becomes prominent for larger lattice volumes than presently used.
\end{abstract}

\pacs{12.38.-t,12.38.Cy,12.39.Fe,12.38.Gc,14.20.-c}

\keywords{Chiral extrapolation, chiral symmetry, Lattice QCD}

\maketitle
\tableofcontents

\newpage
\section{Introduction}
\label{sec:Intro}
The axial-vector form factor of the nucleon, and in particular its value at zero momentum transfer, the axial charge, is a quantity of fundamental interest in hadronic physics. It is central to, for instance, $\beta$ decay and neutrino-nucleon scattering. Next to the nucleon and isobar masses, it is therefore an important testing ground for our understanding of non-perturbative QCD in the framework of chiral perturbation theory. In recent years, good-quality QCD lattice data have become available, not only for the nucleon and isobar masses, but also for the axial-vector form factor of the nucleon.

Chiral perturbation theory is the tool of choice to study the pion and momentum dependence of hadronic quantities. Unfortunately, previous work within flavor-SU(3) chiral perturbation theory has shown serious convergence problems of the chiral expansion. Within flavor-SU(2) chiral perturbation theory, several works have addressed the nucleon and isobar masses and the axial-vector form factor,
by using several expansion schemes. The axial-vector form factor has been calculated to one-loop level in heavy-baryon chiral perturbation theory~\cite{Bernard:1993bq, Bernard:1998gv, Fearing:1997dp} and in relativistic (or covariant) baryon chiral perturbation theory~\cite{Schindler:2006it, Fuchs:2003qc, Chen:2012nx, Ando:2006xy} with different renormalization schemes. The $\Delta(1232)$ isobar was included in Refs.~\cite{Hemmert:2003cb,Procura:2006gq,Yao:2017fym, Ellis:1997kc}. 
%\textbf{(***see also discussion in introduction of \cite{Yao:2017fym} and references therein)}

Motivated by the recent progress in lattice QCD \cite{Capitani:2017qpc,Alexandrou:2017hac,Bali:2018qus}, we develop here a novel scheme based, for convenience, on the relativistic flavor-SU(2) chiral Lagrangian and apply it to the nucleon and isobar masses and the axial-vector form factor of the nucleon. It allows us to explore in an important test case the use of on-shell hadron masses in a chiral approach to systems of pions, nucleons, and isobars. This has implications for an analysis of QCD lattice data of the nucleon axial charge and radius and the pion-nucleon and the pion-isobar sigma terms.

The common challenge faced by treatments of masses and form factors with chiral perturbation theory is the occurrence of power-counting violating terms, with $m_\pi \sim \Delta \sim$ small momenta ($\Delta$ is the isobar-nucleon mass difference), in the chiral expansion \cite{Lutz:2018cqo}. It is not clear how to deal with such terms in other approaches to relativistic baryon chiral perturbation theory. In our scheme, we have to deal with power-counting violating terms in the presence of on-shell masses. The solution lies in considering the chiral Ward identities analyzed in terms of the Passarino-Veltman reduction~\cite{Passarino:1978jh} scheme of the one-loop integrals. We show that we are able to  renormalize the one-loop amplitudes in terms of subtracted Passarino-Veltman integrals, where we drop scalar integrals that involve only baryons, and we need subtractions only in tadpole and bubble terms. An additional challenge occurs in the chiral domain, where we need a dimensional counting with $\Delta \sim M \sim Q^0$. We find that a further class of power-counting violating terms arises, and, therefore, additional subtractions are needed. In this way, we arrive at consistently renormalized amplitudes that can be compared to the QCD lattice data. 

We have organized our paper as follows. In Section~\ref{sec:Lagr} we define the flavor-SU(2) chiral Lagrangians for the nucleon and isobar fields as the starting point of our development. In Section~\ref{sec:Mass} we derive the nucleon and isobar masses at one-loop level. We discuss the Passarino-Veltman framework and our resulting power-counting and renormalization scheme. In Section~\ref{sec:gA} we extend our approach to the axial-vector form factor of the nucleon. Next, in Section~\ref{sec:LQCD}, we present the results of the fits of our expressions to the available QCD lattice data for the nucleon and isobar masses and for the axial-vector form factor. We show that the use of on-shell masses in the one-loop expressions results in non-analytic behavior of the masses \cite{Semke:2006hd, Guo:2019nyp} and form factor as function of the pion mass and its dependence on the lattice size. We discuss the quality of the fit and the resulting values of the parameters. In Section~\ref{sec:Summary} we summarize our findings and conclude with an outlook.

%\newpage
\section{The SU(2) chiral Lagrangians with baryon fields}
\label{sec:Lagr}
%\subsection{}
%\subsubsection{}
We consider the flavor-SU(2) chiral Lagrangian density with nucleon and isobar degrees of freedom \cite{Bernard:2005fy}. We focus on the strict isospin limit with degenerate up- and down-quark masses $m_u=m_d\equiv m$. The isospin-doublet nucleon field is $N^t=(p,n)$. The isospin-triplet pion fields $\vec \pi$  enter via the SU(2) matrix $ \Phi = \vec \tau\, \cdot \,\vec\pi$.
For the nucleon the relevant terms are
\begin{eqnarray}
&& \mathcal{L}_{N}=\bar{N}\big(i\,\slashed{D}-M\big)\,N
+2\,\zeta_N\,\bar{N}\,\chi_+\,\big(i\,\slashed{D}-M\big)\,N
+ 4\,b_\chi\,\bar{N}\, \chi_{+}\,N  +4\,c_\chi\, \bar{N}\,\chi_+^2\,N 
\nonumber\\
&& \quad 
+\, g_A\, \bar{N}\,\gamma^{\mu}\,\gamma_5\,i \,U_{\mu} \,N + 4\,g_\chi\, \bar{N}\,\gamma^{\mu}\,\gamma_5\,\chi_+ \,i\, U_{\mu}\,N
+g_R/2\,\bar{N}\,\gamma^{\mu}\,\gamma_5\,[D^\nu\,,\,F_{\mu\nu}^-]\,N
\nonumber\\
&& \quad 
-\,4\,g_{S}\,\bar{N}\,U_{\mu}\,U^{\mu}\,N 
-\,g_{T}\,\bar{N}\,i\,\sigma^{\mu\nu} \,\big[U_{\mu}\,,\,\,U_{\nu}\big]\, N
 -\,g_{V}\,\Big(\bar{N}\,i\,\gamma^{\mu} \,\{U_{\mu},\,U_{\nu}\}\, D^{\nu} N 
 +{\textrm h.c.}\Big) \,,
\label{def-LN}
\end{eqnarray}
where the nucleon mass parameter is denoted by $M$, and
\begin{eqnarray}
&& U_{\mu}={\textstyle{1\over 2}}\,u^\dagger \,
\Big( (\partial_{\mu}\,e^{i\,\Phi/f})-
\big\{i\,a_{\mu},\,e^{i\,\Phi/f} \big\}\Big) \,u^\dagger \,,\qquad 
\nonumber\\
&&\Gamma_\mu ={\textstyle{1\over 2}}\,u^\dagger \,\big[\partial_\mu -i\, a_\mu \big] \,u
+{\textstyle{1\over 2}}\, u \,
\big[\partial_\mu +i\,a_\mu\big] \,u^\dagger\,,  
\qquad 
F_{\mu\nu}^\pm=u^\dagger\,F^R_{\mu\nu}\,u\, \pm \,u \,F^L_{\mu\nu}\,u^\dagger\,,
\nonumber\\
&&  D_\mu \, N \;\,= \partial_\mu N +  \Gamma_{\mu}\, N  \,, \qquad \qquad 
 u = e^{i\,\Phi /(2\,f)}\,, \qquad \,
\chi_+ = 2\,B_0\,m\, \cos \big(\Phi/f \big) \,,
\nonumber\\
&&F^R_{\mu\nu}=\partial_\mu\,a_\nu-\partial_\nu\,a_\mu-i\,[a_\mu\,,\,a_\nu]\,, \,
  \qquad F^L_{\mu\nu}=-(\partial_\mu\,a_\nu-\partial_\nu\,a_\mu)-i\,[a_\mu\,,\,a_\nu]\,.
\label{def-notation}
 \end{eqnarray}
 
In the presence of the isobar field $\Delta_\mu$ additional terms are required, {\it viz.}
\begin{eqnarray}
&& \mathcal{L}_{\Delta}= - \,\tr \Big[\bar{\Delta}_{\mu}\cdot\big((i\slashed{D}-
(M+\Delta))g^{\mu\nu}
-i\,(\gamma^{\mu}\,D^{\nu}+\gamma^{\nu}\,D^{\mu})+
\gamma^{\mu}(i\,\slashed{D}+(M+\Delta))\,\gamma^{\nu}\big)\,
\Delta_{\nu}\Big]
\nonumber\\
&& \qquad 
- \,2\,\zeta_\Delta\,
\tr\Big[\bar{\Delta}_{\mu}\cdot\big(i\,\slashed{D}-(M+\Delta)\big)\,
\Delta^{\mu}\,\chi_+\,\Big]
-\, 4\,d_\chi
 \,\tr \, \Big[\bar{\Delta}_{\mu}\cdot \Delta^{\mu}\,\chi_{+}\,\Big]
 -4\,e_\chi\,\tr\,\Big[ \bar{\Delta}_{\mu}\cdot\Delta^{\mu}\,\chi_+^2\Big]\,
\nonumber\\
&& \qquad 
+\,h_A \, \tr\Big[ \big(\bar{\Delta}_{\mu}\cdot
\gamma_{5}\,\gamma_{\nu}\,\Delta^{\mu} \big)\,i\, U^{\nu}\Big]
-4\, h_\chi \, \tr\Big[ \big(\bar{\Delta}_{\mu}\cdot
\gamma_{5}\,\gamma_{\nu}\,\Delta^{\mu} \big)\,i\, \chi_+\,U^{\nu}\Big]
\nonumber\\ 
&&\qquad +\,4\,h_{S}^{(1)}\,
 \tr\big[\bar{\Delta}_{\mu}\cdot\Delta^{\mu}\, U_{\nu}\,U^{\nu} \big]
    +4 \,h_{S}^{(2)}\,\big(\bar{\Delta}_{\mu}\cdot U^{\nu}\big)
  \,\big(U_{\nu}\cdot \Delta^{\mu}\big) 
   \nonumber\\
 &&\qquad +\,2\,h_{S}^{(3)}\,
 \tr\big[\bar{\Delta}_{\mu}\cdot\Delta^{\nu}\, \big\{U^{\mu}\,,\,U_{\nu}\big\} \big]
    +2 \,h_{S}^{(4)}\,\Big( \big(\bar{\Delta}_{\mu}\cdot U^{\mu}\big)
  \,\big(U_{\nu}\cdot \Delta^{\nu}\big)
  +\big(\bar{\Delta}_{\mu}\cdot U_{\nu}\big) \,\big(U^{\mu}\cdot \Delta^{\nu}\big)\Big)
  \nonumber\\
&&\qquad
 + \,h_T \,
\tr\Big[ \bar{\Delta}_{\lambda}\cdot i\sigma_{\mu\nu}\Delta^{\lambda}\,
\big[U^{\mu}\,,\,U^{\nu}\big] \Big]
\nonumber\\ 
  &&\qquad +\,h_{V}^{(1)}\,
 \tr\big[\big(\bar{\Delta}_{\lambda}\cdot i\, \gamma^{\mu}\,D^{\nu}\Delta^{\lambda}\big)\,
 \big\{U_{\mu}\,,\,U_{\nu}\big\} \big] +\textrm{h.c.}
 \nonumber\\ 
 &&\qquad + \,h_{V}^{(2)}\,\big(\big(\bar{\Delta}_{\lambda}\cdot U_{\mu}\big)\,i\,\gamma^{\mu}\,
  \,\big(U_{\nu}\cdot D^{\nu} \Delta^{\lambda}\big)
  +\big(\bar{\Delta}_{\lambda}\cdot U_{\nu}\big)\,i\,\gamma^{\mu}\,
  \,\big(U_{\mu}\cdot D^{\nu} \Delta^{\lambda}\big)\big)  +\textrm{h.c.}
   \nonumber\\
&& \qquad+\, f_S\,\Big(\bar{\Delta}_{\mu}\cdot i \,U^{\mu}\,N+ \textrm {h.c.}\Big)
-4\, f_\chi\,\Big(\bar{\Delta}_{\mu}\cdot i \,U^{\mu}\,\chi_+\,N+ \textrm {h.c.}\Big)
\nonumber\\
 &&\qquad
-\, f_{A}^{(1)}\,\tr\big[\big(\bar{\Delta}^{\mu}\cdot \,\gamma^{\nu}\,\gamma_5\,N\big)
 \,\big\{U_{\mu}\,,\,U_{\nu}\big\}\big]+\textrm {h.c.}
 \nonumber\\
  &&\qquad 
 -\, f_{A}^{(2)}\,\tr\big[\big(\bar{\Delta}^{\mu}\cdot \,\gamma^{\nu}\,\gamma_5\,N\big)
 \,\big[U_{\mu}\,,\,U_{\nu}\big]\big]+\textrm {h.c.}
 \nonumber\\
 &&\qquad 
 -\, f_{A}^{(3)}\,\Big(\big(\bar{\Delta}^{\mu}\cdot U_{\nu}\big)
 \,\gamma^{\nu}\,\gamma_5 \,U_{\mu}\, N
 +\big(\bar{\Delta}^{\mu}\cdot U_{\mu}\big)
 \,\gamma^{\nu}\,\gamma_5 \,U_{\nu}\,N\Big)+\textrm {h.c.}
 \nonumber\\
&&\qquad  -\, f_{A}^{(4)}\,\Big(\big(\bar{\Delta}^{\mu}\cdot U_{\nu}\big)
 \,\gamma^{\nu}\,\gamma_5 \,U_{\mu}\, N
 -\big(\bar{\Delta}^{\mu}\cdot U_{\mu}\big)
 \,\gamma^{\nu}\,\gamma_5 \,U_{\nu}\,N\Big)+\textrm {h.c.},
 \label{def-LDelta}
\end{eqnarray}
where the isobar mass parameter is denoted by $M+\Delta$, and
\begin{eqnarray}
&& \Delta_{\mu}^{111}=\Delta_{\mu}^{++}\,,\qquad \; \;
  \Delta_{\mu}^{112}=\Delta_{\mu}^{+}/\sqrt{3}\,,\qquad \; \;
   \Delta_{\mu}^{122}=\Delta_{\mu}^{0}/\sqrt{3}\,,\qquad \; \;	   
   \Delta_{\mu}^{222}=\Delta_{\mu}^{-}\,,
\nonumber\\
&& (\Phi\cdot\Delta_{\mu})^a=\epsilon_{kl3}\,\Phi_n^l\,\Delta_{\mu}^{kna}, \hspace{1cm}
 (\bar{\Delta}^{\mu}\cdot\Phi)_b=\epsilon^{kl3}\,\bar{\Delta}^{\mu}_{knb}\,\Phi^n_l, \hspace{1cm}
 (\bar{\Delta}^{\mu}\cdot\Delta_{\mu})^a_b=\bar{\Delta}^{\mu}_{bcd}\,\Delta_{\mu}^{acd}\,,
 \nonumber\\
&&  (\bar{N}\cdot\Delta_{\mu})_b^a=\epsilon_{k3b}\,\bar{N}_n\Delta_{\mu}^{kna}, \hspace{1cm}
 (\bar{\Delta}^{\mu}\cdot N)_b^a=\epsilon^{k3a}\bar{\Delta}^{\mu}_{knb}\,N^n,
 \nonumber\\
&&  (D_\mu \,\Delta_\nu)^{abc} = \partial_\mu \Delta_\nu^{abc} + \Gamma^{a}_{d,\mu}\,\Delta_\nu^{dbc} 
+ \Gamma^{b}_{d,\mu}\,\Delta_\nu^{adc}+  \Gamma^{c}_{d,\mu}\,\Delta_\nu^{abd}\, .
\label{def-notation-Delta}
 \end{eqnarray} 
 
The Lagrangian densities in Eqs. (\ref{def-LN}) and (\ref{def-LDelta}) contain a number of coupling constants or ``low-energy constants'' (LECs). At leading order in large-$N_c$ they satisfy \cite{Lutz:2010se, Semke:2011ez}
 \begin{eqnarray}\label{LargeNcSU2}
h_A & = & 9g_A-6f_S \ , \hspace{1cm}
 g_V = h_V^{(1)}+\frac{4}{3}h_V^{(2)} \ , \nonumber \\
 g_S & = & h_S^{(1)}+\frac{4}{3}h_S^{(2)}+\frac{4}{9}h_S^{(4)} \ , \hspace{1cm}
 h_S^{(3)} = -h_S^{(4)} \ , \nonumber\\
 f_1^{(A)} & = & 0 \ , \hspace{1cm} f_2^{(A)} = \frac{4}{3}h_T \ , \hspace{1cm} f_4^{(A)}=\frac{10}{3}h_T-6g_T \ , 
 \nonumber \\
 b_{\chi} & = & d_{\chi} \ , \hspace{1cm}  c_\chi = e_\chi \ , \hspace{1cm} \zeta_{N}=\zeta_{\Delta} \ .
\end{eqnarray}

\section{Quark-mass dependence of nucleon and isobar mass}
\label{sec:Mass}
Given the chiral Lagrangian densities in Eqs. (\ref{def-LN}) and (\ref{def-LDelta}) it is straightforward to derive its implications for the nucleon and isobar mass at the one-loop level in dimensional regularization \cite{Gasser:1987rb,Semke:2005sn}. There are various schemes how to deal with the power-counting violating contributions  \cite{Becher:1999he,Gegelia:1999gf,Lutz:1999yr}. Here we follow a framework based on the Passarino-Veltman reduction 
\cite{Passarino:1978jh}. It was argued in Ref.
\cite{Semke:2005sn} that power-counting violating terms come with scalar loop integrals only, which depend on the 
renormalization scale $\mu$. In turn it suffices to set up a suitable subtraction scheme for the scalar tadpole and bubble loop integrals. Such a program was developed for the flavor-SU(3) baryon masses in Refs. \cite{Semke:2005sn,Lutz:2014oxa,Lutz:2018cqo}.  Adapted to our flavor-SU(2) case, the results are
\allowdisplaybreaks[1] 
\begin{eqnarray}
   && M_N=M +\bar{\Sigma}_N(M_N)\,, \qquad \qquad \qquad M_\Delta=M+ \Delta + \Re \,\bar{\Sigma}_\Delta(M_\Delta +i\,\epsilon)\,,
   \nonumber\\
   &&\bar{\Sigma}_{N}=-8\,b_\chi\,B_0\, m -4\,c_\chi \,m_{\pi}^4
   + \frac{3}{f^2}\,\Big( -(\bar g_S+M\,\bar g_V/4)\,m_{\pi}^2   +2\,b_\chi\,m_{\pi}^2\Big)\,\bar{I}_{\pi}
   \nonumber\\
   &&\qquad-2\,\zeta_{N}\,m_{\pi}^2\, (M_N-M)
   +\bar{\Sigma}_N^{\rm bubble}/Z_N \ ,
    \nonumber\\
   &&\bar{\Sigma}_{\Delta}=-8\,d_\chi\,B_0\, m -4\,e_\chi \,m_{\pi}^4
     + \frac{3}{f^2}\,\Big( -(\bar h_S+(M+\Delta)\,\bar h_V/4)\,m_{\pi}^2   +2\,d_\chi\,m_{\pi}^2\Big)\,\bar{I}_{\pi}
     \nonumber\\
     &&\qquad-2\,\zeta_{\Delta}\,m_{\pi}^2\, (M_\Delta-(M+\Delta))
   +\bar{\Sigma}_{\Delta}^{\rm{bubble}}/Z_\Delta \ ,
 % \nonumber\\ \nonumber\\
    \label{def-sigma}
 \end{eqnarray}
 with $m_\pi$ the pion mass, $Z_N$ and $Z_\Delta$  the wave-function renormlization factors of the baryons, and
 \begin{eqnarray}
   &&\quad\bar{g}_S=g_S-\frac{(4\,M^2+\Delta\, M-\Delta^2)}{18\,M\,(M+\Delta)^2}\,f_S^2, \hspace{1cm}
   \bar{g}_V=g_V-\frac{1}{9\,(M+\Delta)^2}\,f_S^2\,,
   \nonumber\\
   &&\quad\bar{h}_S=h_S+\frac{4\,M^2+5\,\Delta\, M+2\,\Delta^2}{72\,(M+\Delta)^3}\,f_S^2,\hspace{1cm}
   h_S=h_{S}^{(1)}+\frac{2}{3}h_{S}^{(2)}+\frac{1}{4}h_{S}^{(3)}+\frac{1}{6}h_{S}^{(4)},
   \nonumber\\
   &&\quad
   \bar{h}_V=h_V-\frac{1}{(M+\Delta)^2}
   \Big(\frac{5}{162}\,h_A^2+\frac{1}{36}\,f_S^2\Big), \hspace{1cm}
  h_V= h_{V}^{(1)}+\frac{2}{3}h_{V}^{(2)}\,.
\end{eqnarray}
All quantities in Eq. (\ref{def-sigma}) are expressed in terms of renormalized scalar loop functions, a 
tadpole $\bar I_\pi$, and a bubble function $\bar I_{\pi R}(M_B)$ with $R,B \in \{N, \Delta\}$. 
We identify the mass of the isobar $M_\Delta$ in a quasi-particle approach where its 
width is determined by $ \Im \,\bar{\Sigma}_\Delta(M_\Delta + i\,\epsilon)$. 

The loop functions take the form
\begin {eqnarray}
&& \bar I_\pi = \frac{m_\pi^2}{16\,\pi^2}\,\log\Big[ \frac{m_\pi^2}{\mu^2}\Big] \,, \qquad \qquad \qquad \qquad 
\bar I^{}_{\pi| R} = \frac{m_\pi^2}{16\,\pi^2}\,\log \Big[\frac{m_\pi^2}{M_R^2}\Big]\,,
\nonumber\\
&& \bar{I}_{\pi R}\big(M_B\big)=
 \frac{1}{16\pi^2} \,\Bigg[\gamma_B^R-\Big(\frac{1}{2}+\frac{m_\pi ^2-M_R^2}{2\,M_B^2}\Big)
 \log\,\Big(\frac{m_\pi^2}{M_R^2}\Big)
 \nonumber\\ 
&& \qquad +\,\frac{p_{\pi R}}{M_B}\Big(\log\,\Big[1-\frac{M_B^2-2\,p_{\pi R}\,M_B}{m_\pi ^2+M_R^2}\Big]-
 \log\,\Big[1-\frac{M_B^2+2\,p_{\pi R}\,M_B}{m_\pi ^2+M_R^2}\Big]\Big)\Bigg]\,, \label{def-scalar}
%\nonumber\\
\end{eqnarray}
where
\begin{eqnarray}
 p_{\pi R}^2 & = & \frac{M_B^2}{4}-\frac{M_R^2+m_{\pi}^2}{2}+\frac{(M_R^2-m_\pi^2)^2}{4\,M_B^2} \ , \\%\qquad \quad \!\! 
 \gamma_B^R & = & -\lim_{m\to 0}\frac{M_R^2-M_B^2}{M_B^2} \, \log\,\Big|\frac{M_R^2-M_B^2}{M_R^2} \Big|\, .
% \nonumber
\end{eqnarray}
The tadpole $\bar I_\pi $ depends on 
the renormalization scale $\mu$ of dimensional regularization, while the renormalized bubble $\bar I_{\pi R}(M_B)$ in our scheme does not. The subtraction term $\gamma_B^R$ illustrates the necessity of 
a renormalization scheme that leads to results that are in accordance with the expectation of dimensional power-counting rules. In the chiral domain with $m_\pi < M_\Delta - M_N \sim Q^0$ one expects $\bar I_{\pi N}(M_N) \sim Q$, but $\bar I_{\pi \Delta}(M_N) \sim Q^2$, with $Q$ denoting the chiral small scale $Q \sim m_\pi$. Given the form of our renormalized bubble function such a behavior is ensured. Clearly, this is not the case when $\gamma_B^R = 0$.

Matters turn considerably more complex once we start to leave the chiral domain and consider 
$m_\pi \sim M_\Delta -M_N$, or even $M_\Delta - M_N < m_\pi < 4\pi f$. In order to avoid a proliferation of counting schemes we follow here a pragmatic path, where we simply keep all model-independent parts of the one-loop expressions. The bubble-loop contribution to the nucleon mass is given by (see Eq. (31) of Ref. \cite{Lutz:2018cqo})
  \begin{eqnarray}
  &&f^2\, \bar{\Sigma}_{B=N}^{\rm bubble}=\frac{3}{4}\,g_A^2\,
   \Bigg\{\frac{M_N^2-M_B^2}{2\,M_B}\,\bar{I}^{}_{\pi | N}-
   \frac{(M_B+M_N)^2}{E_N+M_N}\,p_{\pi N}^2\,\bar{I}_{\pi N}\Bigg\} \,,
\nonumber\\
&& \qquad  \;+\,f_S^2\,
   \Bigg\{\frac{(M_{\Delta}+M_B)^2}{12\,M_B\,M_{\Delta}^2}
   \big(M_{\Delta}^2-M_B^2\big)\,\bar{I}^{}_{\pi | \Delta}
 -\,\frac{2\,M_B^2}{3\,M_{\Delta}^2}\,(E_{\Delta}+M_{\Delta})\,p_{\pi \Delta}^2\,\bar{I}_{\pi \Delta}
  +\frac{4}{3}\,\alpha_{\pi\Delta}^{(B)}\Bigg\}\,,
\label{def-N-bubble}
\end{eqnarray}
with
\begin{eqnarray}
&&\alpha_{\pi\Delta}^{(B=N)}=\frac{\alpha_1\Delta^2}{16\pi^2}\,
    \Big(M_{\Delta}-\Big(1+\frac{\Delta}{M}\Big)\,M_B\Big) \,
    \Big(\frac{\Delta\partial}{\partial\Delta}+1\Big)\,\gamma_1 + \frac{\Delta \,m_{\pi}^2}{16\,\pi^2}\,\alpha_1\,\gamma_2 \,,
\nonumber\\
&& E_{\Delta}^2=p_{\pi \Delta}^2+M_\Delta^2 \,,   \qquad 
  \qquad  \gamma_1=\frac{2\,M+\Delta}{M}
    \log \Big[\frac{\Delta(2\,M+\Delta)}{(M+\Delta)^2}\Big]\,, \quad 
\\
&&\alpha_1=\frac{(2\,M+\Delta)^4}{16\,M^2\,(M+\Delta)^2} \,,\quad \;\, \gamma_2=-\frac{2\,M^2+2\,\Delta M+\Delta^2}{M\,(2\,M+\Delta)}
    \log\Big[\frac{\Delta(2\,M+\Delta)}{(M+\Delta)^2}\Big]
    -\frac{M}{2\,M+\Delta} \,.
\nonumber
\end{eqnarray}
Consider first the term proportional to $\bar I_{\pi \Delta}$. The pre-factor 
$p^2_{\pi \Delta}$ is model independent, since it is the relativistic phase-space factor 
describing the $N \to \pi\,\Delta$ decay process, which is accessible for some unphysical parameter choices of the chiral Lagrangian. At $M_N > M_\Delta + m_\pi$ the scalar bubble loop function 
$\bar I_{\pi \Delta}$ turns complex. The full pre-factor cannot change upon the consideration of higher-loop effects. The only effect expected is that the bare coupling constant $f_S$ is replaced by its on-shell physical value, which then of course may show some quark-mass dependence. The role of the other term proportional to $\bar I_{\pi | \Delta}$ is more subtle. In fact, there is important cross talk to the $\bar I_{\pi \Delta}$ term \cite{Lutz:2018cqo}. The reason is that a term $\sim \Delta\,m_\pi^2\,\log m_\pi^2$ cannot be absorbed into any of the other non-bubble term contributions in Eq. (\ref{def-sigma}). 
Similar arguments can be put forward in favor of the model-independent nature of the terms proportional to $\bar I_{\pi N}$ and $\bar I_{\pi | N}$. 

There is yet the subtraction term $\alpha^{(N)}_{\pi \Delta}$ to be discussed. The purpose of the latter is twofold. 
First, it ensures that the baryon wave-function renormalization factor becomes unity in the chiral limit, {\em viz.}
\begin{eqnarray}
 Z_B=\big(1 - 2\,m_\pi^2\,\zeta_B + \frac{\partial}{\partial M_B}\bar{\Sigma}_B^{\rm bubble}\big)  \bigg|_{m \to 0} \to 1 \, .
\end{eqnarray}
Second, it is required to protect the  non-analytic $g_A^2 \,m_\pi^3$ contribution to the nucleon mass. A corresponding term $h_A^2 \, m_\pi^3$ for the isobar mass would generate 
a contribution proportional to $h_A^2\,g_A^2\,\Delta^2\,m_\pi^3$, which must be cancelled by higher-loop effects \cite{Pandharipande:2005sx,Long:2009wq,Lutz:2018cqo}. We run into this issue only because we wish to keep the 
on-shell hadron masses inside the loop functions. 

We now turn to the bubble-loop contributions to the isobar mass,
\begin{eqnarray}
 && f^2\, \bar{\Sigma}_{B=\Delta}^{\rm bubble}=\frac{f_S^2}{2}\,
   \Bigg\{\frac{(M_{N}+M_B)^2}{24\,M_B^3}\big(M_{N}^2-M_B^2\big)\bar{I}^{}_{\pi | N}-
   \frac{1}{3}(E_{N}+M_{N})\, p_{\pi N}^2\bar{I}_{\pi N}+
   \frac{2}{3}\,\alpha_{\pi N}^{(B)}\Bigg\}
   \nonumber\\
&& \qquad \;+\,\frac{5}{12}\,h_A^2\,
   \Bigg\{\Big(-\frac{M_{\Delta}^2+M_B^2}{18\,M_B^2\,M_{\Delta}}
   +\frac{M_{\Delta}^4+M_B^4+12\,M_{\Delta}^2\,M_B^2}{36\,M_B^3\,M_{\Delta}^2}\Big)
   \big(M_{\Delta}^2-M_B^2\big)\bar{I}^{}_{\pi | \Delta}
\nonumber\\ 
&&  \hspace{2cm} -\frac{(M_B+M_{\Delta})^2}{9\,M_{\Delta}^2}
  \frac{2\,E_{\Delta}(E_{\Delta}-M_{\Delta})+5\,M_{\Delta}^2}{E_{\Delta}+M_{\Delta}}\,
  p_{\pi \Delta}^2\,\bar{I}_{\pi \Delta} 
 \Bigg\}\,,
\label{def-Delta-bubble}
\end{eqnarray}
with
\begin{eqnarray}
&&\alpha_{\pi N}^{(B=\Delta)}=\frac{\beta_1\,\Delta^2}{16\,\pi^2}
    \Big(M_{B}-\Big(1 + \frac{\Delta}{M}\Big)\, M_N \Big)
    \Big(\frac{\Delta\partial}{\partial\Delta}+1\Big)\,\delta_1 
     +\,\frac{\Delta \,m_{\pi}^2}{16\,\pi^2}\,\beta_1\,\delta_2 \,,
\nonumber\\
&& E_{N}^2=p_{\pi N}^2+M_N^2 \,, 
   \qquad
  \qquad \!\delta_1=-\frac{M\,(2\,M+\Delta)}{(M+\Delta)^2}
    \log\,\Big[\frac{\Delta(2\,M+\Delta)}{M^2}\Big]\,, \qquad 
\\
&& \beta_1=\frac{(2\,M+\Delta)^4}{16\,M\,(M+\Delta)^3}\,,\qquad
\delta_2=\frac{M}{2\,M+\Delta}+\frac{M\,(2\,M^2+2\,\Delta M+\Delta^2)}{(2\,M+\Delta)(M+\Delta)^2}
    \log\,\Big[\frac{\Delta\,(2\,M+\Delta)}{M^2}\Big]\,,
\nonumber
\end{eqnarray}
where we again argue that the terms shown are model independent. The subtraction term $\alpha_{\pi N}^{(\Delta)}$ is instrumental to avoid the consideration of explicit contributions from two-loop effects. By construction the wave-function renormalization factor of the isobar approaches unity in the chiral limit. 

It is worth pointing out another subtle issue. 
The expressions in Eqs. (\ref{def-N-bubble}) and (\ref{def-Delta-bubble}) show a non-trivial dependence on the ratio $\Delta/M$. While 
an expansion in powers of that ratio is formally convergent, it is advantageous not to expand, since 
the convergence pattern is rather poor. Any significant result would require more terms than one would consider even at the N$^3$LO level. Thus it is better  to keep the unexpanded form.

\section{Axial charge and radius of the  nucleon}
\label{sec:gA}
Consider the matrix element of the axial-vector current in the nucleon state  \cite{Gasser:1987rb},
\begin{eqnarray}
 \bra{N(\bar{p})}\,A_{i}^{\mu}(0)\,\ket{N(p)}=
 \bar{u}_{N}(\bar{p})\,\Big(\gamma^{\mu}\,\gamma_5 \,\frac{\tau_i}{2}\, G_A(q^2)+
 \gamma_5 \,\frac{q^{\mu}}{2\,M_N} \,\frac{\tau_i}{2}\, G_P(q^2)\Big)\,u_N(p),
 \label{Axialcurrent}
\end{eqnarray}
with $q = \bar p -p$. Given the chiral Lagrangian densities of Eqs. (\ref{def-LN}) and (\ref{def-LDelta}) it is straightforward to derive its tree-level and one-loop contributions to the two form factors $G_{A}(q^2)$ and $G_{T}(q^2)$. In this work we focus on the axial term $G_A(q^2)$, which defines the axial charge $G_A(0)$ and radius $\langle r_A^2\rangle \equiv 6\,G_A'(0)/G_A(0)$ of the nucleon. In the chiral limit at $m=0$ it holds that
$G_A(0)\to g_A$.

In the Passarino-Veltman reduction scheme \cite{Passarino:1978jh} we find for
the one-loop contributions to the axial-vector form factor
 \begin {eqnarray}
 && G_A(q^2) = g_A \, Z_N +4\, g_\chi\, m_{\pi}^2 + g_R \, q^2 + K_\pi \,\bar I_\pi + \sum_{R = N, \Delta}\,K_ {\pi R}\,\bar I_{\pi R}(M_N)
 \nonumber\\
 &&+ \sum_{L,R = N, \Delta}\,K_ {L \pi R}\,\bar I_{L \pi R} (q^2) \,
 + \sum_{L,R = N, \Delta}\,K'_ {L \pi R}\,\frac{\Delta\bar{I}_{L\pi R}}{q^2} \,,
 \label{def-}
 \end{eqnarray}
 where the kinematical functions $K_\pi, K_{\pi R}$, $K_{L\pi R}$,  and $K'_{L\pi R}$ depend on the pion mass $m_\pi$, the momentum transfer $t= q^2$, and the baryon masses $M_N$ and $M_\Delta$. 
While we derived explicit expressions thereof, they are too lengthy to be shown here in full detail. In Appendix A and Appendix B  we document a recursion scheme in terms of which they were derived.  Below we will display the leading-order terms of their chiral decomposition.
  
The scalar tadpole and bubble integrals $\bar I_\pi$ and $\bar I_{\pi R}(M_N)$ we encountered already in the previous section on the baryon-mass evaluations, where also the wave-function renormalization factor for the nucleon $Z_N$ was introduced. We recall that within our renormalization scheme all tadpole terms $\bar I_N$ or $\bar I_\Delta$ that involve a heavy field must be dropped, because they imply power-counting violating contributions to the form factors. As emphasized, such a procedure complies with the chiral Ward identities of QCD, simply because manipulating both sides of a Ward identity in a well-defined manner does not spoil it \cite{Semke:2005sn}. 
 
The evaluation of the axial-vector form factor involves another function of the Passarino-Veltman basis, the scalar triangle function $I_{L\pi R}(t)$. This function is strictly finite and for $R=L=N$ it follows the dimensional counting rules with $I_{N\pi N}(t) \sim Q^0$. In the presence of the isobar degrees of freedom it still holds that $I_{L\pi R}(t) \sim Q^0$ for $L,R \in \{N,\Delta\}$, however, only if we count $\Delta \sim Q \sim m_\pi$. The trouble starts in the chiral domain where $m_\pi < \Delta \sim Q^0$. In this case dimensional counting rules require a different scaling behaviour of $I_{L\pi R}(t)$ for $L \neq N$ or $R \neq N$. In order to restore the proper scaling in that domain we implement a subtraction term in the scalar triangle, {\em viz.}
\begin{eqnarray}
&& \bar I_{L\pi R}(q^2)=\int \frac{d^4l}{(2\pi)^4}
\frac{i}{((l-\bar{p})^2-M_{L}^2)(l^2-m_\pi^2)((l-p)^2-M_R^2)} - \gamma_{L\pi R}\,,
\nonumber\\ \nonumber\\
&&\gamma_{\Delta\pi N}=\gamma_{N\pi\Delta}=
-\frac{1}{16\pi^2\,M^2}\log\Big[\frac{2\,M\,\Delta+\Delta^2}{(M+\Delta)^2}\Big]+\frac{1}{16\pi^2\,(2\,M\,\Delta+\Delta^2)}\log\Big[\frac{(M+\Delta)^2}{M^2}\Big]\,,
\nonumber\\
&&\gamma_{N\pi N}=0\,,
\qquad\qquad \gamma_{\Delta\pi\Delta}=
-\frac{1}{16\pi^2\,M^2}\log\Big[\frac{2\,M\,\Delta+\Delta^2}{(M+\Delta)^2}\Big]\,,
\label{res-gammaLpiR}
\end{eqnarray}
 \begin{eqnarray}
 && \Delta\bar{I}_{L\pi R}=\bar{I}_{L \pi R} (q^2)-\bar{I}_{L \pi R} (0)-q^2\,\gamma'_{L\pi R}\,, 
  \nonumber\\
 &&\gamma'_{N\pi \Delta}
 =\gamma'_{\Delta\pi N}
=-\frac{1}{96\pi^2\, M^4\, \Delta^3\, (2\,M + \Delta)^3}\Bigg(\Delta^3 \,(2\,M + \Delta)^3 \log\Big[\frac{\Delta}{M}\Big] 
    \nonumber\\
    &&\qquad - 2\, (M + \Delta)^2 (4\,M^4 - 2\, M^3 \,\Delta + 3\,M^2 \,\Delta^2  + 4\, M\, \Delta^3 + \Delta^4) \log\Big[\frac{M + \Delta}{M}\Big]  
   \nonumber\\
   &&\qquad +\Delta \,(2\,M + \Delta)\Big(M^2\,(4\,M^2 + 2 \, M\,\Delta + \Delta^2) + \Delta^2 \,(2\, M + \Delta)^2 \log\Big[\frac{2\,M + \Delta}{M}\Big]\Big)\Bigg)\,,
   \nonumber\\
 &&
 %d=\frac{\Delta}{M}\,,\hspace{2cm} 
 \gamma'_{N\pi N}=\gamma'_{\Delta\pi \Delta}=0\,, 
 \label{res-gammaprimeLpiR}
 \end{eqnarray}
with $p^2=\bar p^2=M_N^2$. This is analogous to the subtraction term $\gamma^\Delta_N$ introduced in the scalar bubble function, which leads to $\bar I_{\pi \Delta}(M_N) \sim Q^2$. With Eq. (\ref{res-gammaLpiR}) we obtain 
$ \bar I_{N\pi N}(t)\sim Q^0$, $ \bar I_{N\pi \Delta }(t)\sim Q$, and 
$ \bar I_{\Delta \pi \Delta }(t)\sim Q^2$ in the chiral domain. 
It is crucial to consider these subtractions, since otherwise the explicit evaluation of a class of two-loop diagrams would be needed \cite{Long:2009wq}. In Appendix C we provide some  
more explicit expression for the triangle function that are instrumental in the computation  of the axial charge and radius of the nucleon. 
 
We return to the kinematic functions $K_\pi, K_{\pi R}$, $K_{L\pi R}$, and $K'_{L\pi R}$. One may be tempted to simply keep their form as they come out of our computation scheme in Appendix A and Appendix B. However, we argue that this would lead to inconsistencies. This is immediately clear by looking for instance into the function $K_\pi$. The renormalization scale dependence in $\bar I_\pi$ can be balanced by the LEC $g_\chi$ only if the suitably renormalized term  $K_\pi$ is independent on the quark mass. Within our scheme $\bar I_\pi$ is the exclusive source of such a dependence. Both our bubble and triangle functions do not depend on $\mu$.
 
We conclude that it is crucial to consider a chiral decomposition of such kinematic functions. How to do so in a scheme wherein one keeps on-shell masses $M_N$ and $M_\Delta$ needs development. For this purpose the functions are taken to depend on $t$,  $m_\pi$, $\delta$, and  $M_N$, with
 \begin{eqnarray}
 &&m_\pi^2\sim Q^2\,,\qquad \qquad \delta = M_\Delta - M_N\,(1 + \Delta/M) \sim Q^2 
 \,, \qquad  
 \nonumber\\
&&t\sim Q^2\,,\qquad \quad \qquad M_N \sim Q^0 \,.
\label{def-expansion}
 \end{eqnarray}
This choice implies a well-defined chiral expansion of the form
 \allowdisplaybreaks[1]
 \begin{eqnarray}
 && K_\pi = \frac{-g_A + g_A^3/4}{f^2}+ 
\frac{4\,g_A\,f_S^2}{9\,f^2}\,\alpha_{01} + \frac{20\,h_A\,f_S^2}{81\,f^2} \,\alpha_{02} 
-\frac{20\,f_S\,M_N\,(5\,f_A^{(3)}+f_A^{(4)})}{27\,f^2}\,\frac{\Delta}{M}\,\alpha_{03}
+ {\mathcal O }\big( Q^2\big)\,,
 \nonumber\\
 && K_{\pi N} = \frac{-2\,g_A+g_A^3/4+8/3 \,g_A\,M_N \,(g_S-2\,g_T)}{f^2}\,m_\pi^2 
 \nonumber\\
 &&\qquad+ \frac{2\,g_A\,f_S^2}{3\,f^2}\,\Big\{
 -\frac{5}{6}\frac{\Delta}{M}\, \alpha_{10}\,M_N^2
 -\frac{5}{24} \,\frac{\Delta}{M}\,\alpha_{11}\,t
 - \frac{5}{18}\,\alpha_{12}\,m_\pi^2 
 - \frac{5}{6}\,\alpha_{13}\,M_N\,\delta   \Big\}
+ {\mathcal O} \big(Q^4\big)\,,\qquad 
 \nonumber\\
  &&  K_{\pi \Delta} = 
 \frac{2\,g_A\,f_S^2}{3\,f^2}\,\Big\{ 
 -\frac{5}{6}\,\frac{\Delta}{M}\, \alpha_{20}\,M_N^2
 -\frac{5}{24}\,\frac{\Delta}{M}\,\alpha_{21}\,t
 + \frac{19}{18}\,\alpha_{22}\,m_\pi^2 
 - \frac{5}{6}\,\alpha_{23}\,M_N\,\delta  \Big\} 
 \nonumber\\
&& 
\qquad  -\,\frac{5\,h_A\,f_S^2}{9\,f^2}\,\Big\{ 
\frac{14}{9}\,\frac{\Delta}{M}\, \alpha_{30}\,M_N^2
-\frac{1}{18}\,\frac{\Delta}{M}\,\alpha_{31}\,t 
-\frac{1}{27}\,\alpha_{32}\,m_\pi^2 
+ \frac{14}{9}\,\alpha_{33}\,M_N\, \delta   \Big\}
\nonumber\\
&& \qquad -\, \frac{2\,f_S\, M_N\,\big(5\,f_A^{(3)}+f_A^{(4)}\big)}{3\,f^2}\,\Big\{
-\frac{20}{9}\,\frac{\Delta^2}{M^2}\, \alpha_{40}\,M_N^2  
+ \frac{20}{9} \,\alpha_{42}\, m_\pi^2
-\frac{40}{9}\,\frac{\Delta}{M}\,\alpha_{43}\,M_N\,\delta\Big\}
+\, {\mathcal O} \big(Q^4\big)\,,
 \nonumber\\
 && K_{N\pi N} =  {\mathcal O} \big(Q^4\big) \,,
 \nonumber\\
 && K_{N \pi \Delta} = \frac{2\,g_A\,f_S^2\,M_N^2}{3\,f^2}\,\Big\{ 
 \frac{5}{6}\,\frac{\Delta^2}{M^2}\, \alpha_{50}\,M_N^2 
 + \frac{5}{24}\,\frac{\Delta^2}{M^2}\,\alpha_{51}\,t 
 -2\, \alpha_{52}\,  m_\pi^2 
 +\frac{5}{3} \,\frac{\Delta}{M}\,\alpha_{53}\,M_N\,\delta 
 \Big\} + {\mathcal O} \big(Q^4\big) \,,
  \nonumber\\
  &&K'_{N\pi \Delta }  =\frac{2\,g_A\,f_S^2\,M_N^4}{3\,f^2}\,\Big\{
\frac{2}{3}\,\frac{\Delta^2}{M^2}\, \alpha_{60}\,M_N^2 
- \frac{2}{3} \,\frac{\Delta^2}{M^2}\,\alpha_{62}\, m_\pi^2
+\frac{4}{3}\,\frac{\Delta}{M}\,\alpha_{63}\,M_N\,\delta\Big\} +{\mathcal O} \big(Q^4\big)\,, 
  \nonumber\\
 && K_{\Delta \pi \Delta} =-\frac{5\,h_A\,f_S^2 \,M_N^2}{9\,f^2}\,\Big\{
 -\frac{4}{3}\,\frac{\Delta^2}{M^2}\, \alpha_{70}\,M_N^2
 -\frac{7}{9}\,\frac{\Delta^2}{M^2}\,\alpha_{71}\,t 
 + \frac{4}{3}\,\alpha_{72}\,  m_\pi^2 
 -\frac{8}{3}\, \frac{\Delta}{M}\,\alpha_{73}\,M_N\,\delta \Big\}  
 +{\mathcal O} \big(Q^4\big)  \,,
\nonumber\\
&& K_{\Delta \pi N}  = K_{N \pi \Delta} \,,
  \hspace{2cm}
K'_{\Delta \pi N}  = K'_{N \pi \Delta} \,,
   \hspace{2cm}
K'_{N \pi N}  = K'_{\Delta \pi \Delta} = 0 \,.
\label{K-one-loop}
\end {eqnarray}

\begin{figure}[t]
\includegraphics[width=150mm]{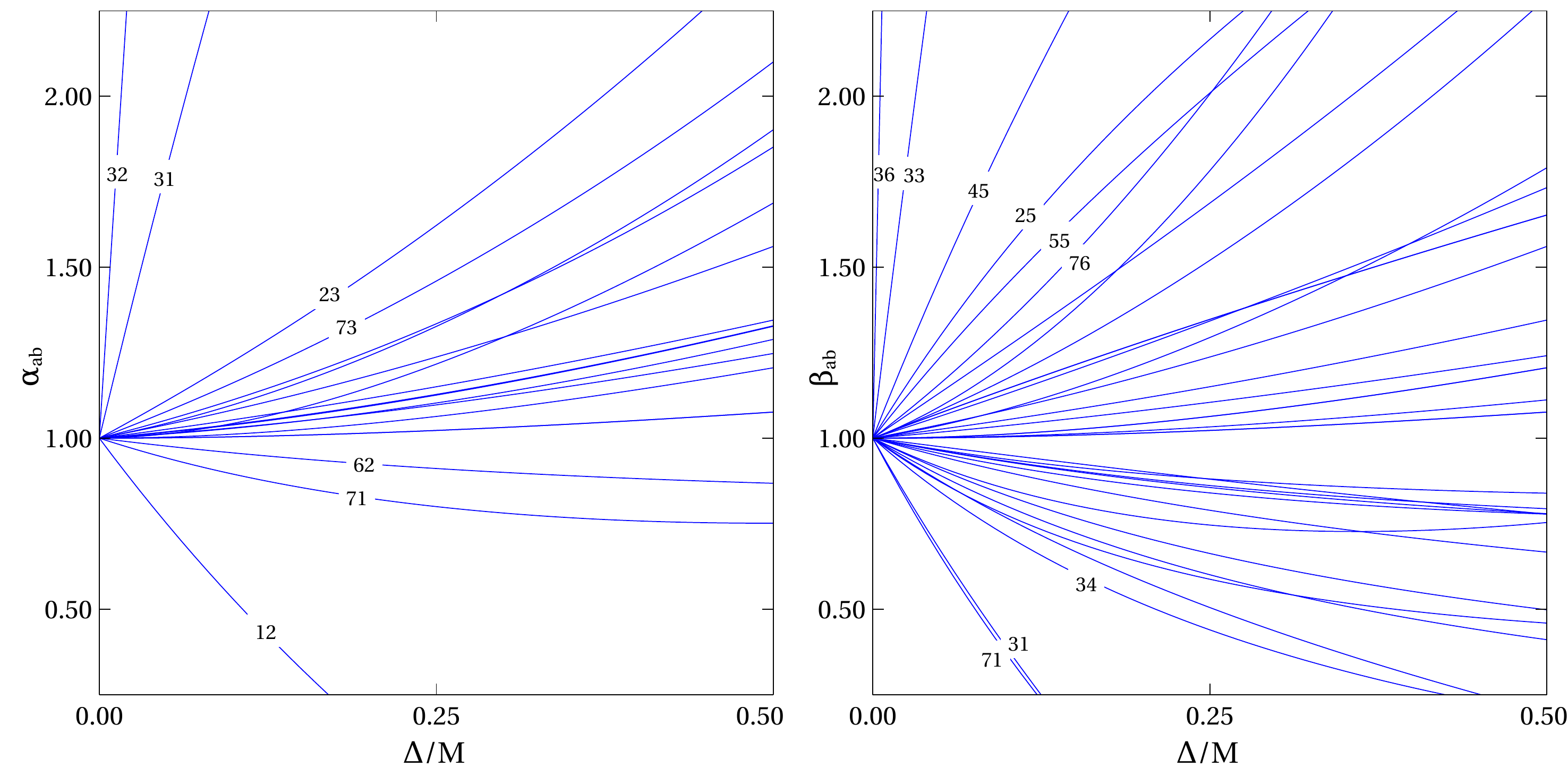}
\caption{All coefficients $\alpha_{ab}$ and $\beta_{ab}$ with $a \neq 0 \neq b$ as function of the ratio $\Delta/M$.}
\label{PlotAlphaBeta}
\end{figure}

The next higher-order contributions to the $K^{(n)}$ coefficients are given by
\begin{eqnarray}
&& K^{(2)}_{\pi} =
-\frac{g_A}{f^2\,M_N}\,\beta_{01}\,\Big(\frac{2}{3}\,g_S+\frac{2}{3}\,g_T+\frac{5}{4}\,M_N\,g_V\Big)\,m_\pi^2
  +\frac{2\,g_A\,f_S^2}{3\,f^2\,M_N^2}\Big(\frac{1}{18}\,\beta_{02}\,m_\pi^2+\frac{1}{9}\,\beta_{03}\,M_N\,\delta\Big)
\nonumber\\
&&\qquad
-\frac{5\,h_A\,f_S^2}{9\,f^2\,M_N^2}\Big(\frac{2}{9}\,\beta_{04}\,t+\frac{133}{108}\,\beta_{05}\,m_\pi^2+\frac{10}{27}\,\beta_{06}\,M_N\,\delta\Big)
\nonumber\\
&&\qquad  -\, \frac{2\,f_S\,\big(5\,f_A^{(3)}+f_A^{(4)}\big)}{3\,f^2\,M_N}
\Big(\frac{17}{18}\,\beta_{07}\,m_\pi^2+\frac{10}{9}\,\beta_{08}\,M_N\,\delta\Big)
\,,
\nonumber\\
  && K^{(4)}_{\pi N} =  -\frac{g_A}{3\, f^2\,
M_N}\Big(g_V\,M_N\,t\,m_\pi^2+2\,\big(g_S+g_T\big)\,m_\pi^4\Big)-\frac{g_A^3}{16\,f^2\,M_N^2}\,t\,m_\pi^2
  \nonumber\\
  &&\qquad +\frac{2\,g_A\,f_S^2}{3\,f^2\,M_N^2}\,\Big\{
-\frac{5}{96}\,\frac{\Delta}{M}\,\beta_{11}\,t^2
+\frac{1}{24}\,\beta_{12}\,t \, m_\pi^2
-\frac{5}{24}\,\beta_{13}\,t\, M_N\, \delta
\nonumber\\
&&\qquad \qquad\qquad\,
-\frac{1}{18}\,\beta_{14}\,m_\pi^4
+\frac{53}{36}\,\beta_{15}\,m_\pi^2\,M_N\,\delta
-\frac{9}{8}\,\frac{\Delta}{M}\,\beta_{16}\,M_N^2\,\delta^2\Big\}
\,,
  \nonumber\\
   &&  K^{(4)}_{\pi \Delta} =
  \frac{2\,g_A\,f_S^2}{3\,f^2\,M_N^2}\,\Big\{
-\frac{5}{96}\,\frac{\Delta}{M}\,\beta_{21}\,t^2
+\frac{1}{24}\,\beta_{22}\,t \, m_\pi^2
-\frac{5}{24}\,\beta_{23}\,t\, M_N\, \delta
\nonumber\\
&&\qquad \qquad\qquad\,
-\frac{11}{18}\,\beta_{24}\,m_\pi^4
+\frac{1}{4}\,\beta_{25}\,m_\pi^2\,M_N\,\delta
-\frac{8}{9}\,\beta_{26}\,M_N^2\,\delta^2\Big\}
  \nonumber\\
&&
\qquad  -\,\frac{5\,h_A\,f_S^2}{9\,f^2\,M_N^2}\,\Big\{
-\frac{1}{72}\,\frac{\Delta}{M}\,\beta_{31}\,t^2
+\frac{41}{36}\,\beta_{32}\,t \, m_\pi^2
-\frac{1}{18}\,\beta_{33}\,t\, M_N\, \delta
\nonumber\\
&&\qquad \qquad\qquad\,
-\frac{19}{36}\,\beta_{34}\,m_\pi^4
-\frac{127}{54}\,\beta_{35}\,m_\pi^2\,M_N\,\delta
+\frac{1}{27}\,\beta_{36}\,M_N^2\,\delta^2\Big\}
\nonumber\\
&& \qquad -\, \frac{2\,f_S\,
\big(5\,f_A^{(3)}+f_A^{(4)}\big)}{3\,f^2\,M_N}\,\Big\{
-\frac{10}{9}\,\beta_{44}\,m_\pi^4
+\frac{4}{9}\,\beta_{45}\,m_\pi^2\,M_N\,\delta
-\frac{20}{9}\,\beta_{46}\,M_N^2\,\delta^2\Big\} \,,
\nonumber\\
&& K^{(4)}_{N\pi N}=-\frac{g_A^3}{4\,f^2}\,m_\pi^4 \,,
\nonumber\\
  && K^{(4)}_{N \pi \Delta} = \frac{2\,g_A\,f_S^2}{3\,f^2}\,\Big\{
\frac{5}{96}\,\frac{\Delta^2}{M^2}\,\beta_{51}\,t^2
-\frac{1}{4}\,\frac{\Delta}{M}\,\beta_{52}\,t \, m_\pi^2
+\frac{5}{12}\,\frac{\Delta}{M}\,\beta_{53}\,t\, M_N\, \delta
\nonumber\\
&&\qquad \qquad\qquad\,
+\frac{7}{6}\,\beta_{54}\,m_\pi^4
-\frac{2}{3}\,\beta_{55}\,m_\pi^2\,M_N\,\delta
+\frac{5}{6}\,\beta_{56}\,M_N^2\,\delta^2\Big\}  \,,
   \nonumber\\
   &&K^{(4)'}_{\Delta \pi N} =\frac{2\,g_A\,f_S^2\,M_N^2}{3\,f^2}\,\Big\{
%0 \beta_{74}\,m\pi^4
-\frac{4}{3} \,\frac{\Delta}{M}\,\beta_{65}\,
m_\pi^2\,M_N\,\delta
+\frac{2}{3}\,\beta_{66}\,M_N^2\,\delta^2\Big\}\,,
   \nonumber\\
  && K^{(4)}_{\Delta \pi \Delta} =-\frac{5\,h_A\,f_S^2}{9\,f^2}\,\Big\{
\frac{1}{36}\,\frac{\Delta^2}{M^2}\,\beta_{71}\,t^2
-\frac{\Delta}{M}\,\beta_{72}\,t \, m_\pi^2
-\frac{14}{9}\,\frac{\Delta}{M}\,\beta_{73}\,t\, M_N\,\delta
\nonumber\\
&&\qquad \qquad\qquad\,
-\frac{8}{9}\,\beta_{74}\,m_\pi^4
+\frac{4}{3}\,\beta_{75}\,m_\pi^2\,M_N\,\delta
-\frac{4}{3}\,\beta_{76}\,M_N^2\,\delta^2\Big\} \,,
   \nonumber\\
&& K^{(4)}_{\Delta \pi N}  = K^{(4)}_{N \pi \Delta} \,,
   \hspace{2cm}
K^{(4)'}_{\Delta \pi N}  = K^{(4)'}_{N \pi \Delta} \,,
    \hspace{2cm}
K^{(4)'}_{N \pi N}  = K^{(4)'}_{\Delta \pi \Delta} = 0 \,,
\label{K-higher-order}
\end{eqnarray}
where the dimensionless coefficients $\alpha_{ab}$ and $\beta_{ab}$ depend on the ratio $\Delta/M$ only. They are normalized with $\alpha_{ab},\beta_{ab} \to 1$ in the limit $\Delta \to 0$. A complete collection of the coefficients $\alpha_{ab}$ can be found in Appendix D. In Figure \ref{PlotAlphaBeta} we plot the coefficients $\alpha_{ab}$ and $\beta_{ab}$ as function of the ratio $\Delta/M$.

\begin{table}[h]
\centering%\scriptsize
\renewcommand{\arraystretch}{1.0}
\begin{tabular}{c |c c c|c |c c c}\hline\hline
$\alpha_{ab}$& $\Delta/M=0.2$ & $\Delta/M=0.3$ & $\Delta/M=0.4$&$\alpha_{ab}$& $\Delta/M=0.2$ & $\Delta/M=0.3$ & $\Delta/M=0.4$ 
\\ \hline
$\alpha_{01}$&1.04& 1.06& 1.09&
$\alpha_{02}$&0.92& 0.92& 0.93\\  %\hline
$\alpha_{03}$&0.96& 0.95& 0.95&&&&
\\ \hline% \hline
$\alpha_{10}$&1.02& 1.03& 1.05&
$\alpha_{20}$&1.23& 1.36& 1.50\\  %\hline
$\alpha_{11}$&1.02& 1.03& 1.05&
$\alpha_{21}$&1.02& 1.03& 1.05\\ %\hline 
$\alpha_{12}$&0.14& -0.18& -0.45&
$\alpha_{22}$&1.07& 1.12& 1.18\\ %\hline
$\alpha_{13}$&1.04& 1.09& 1.14&
$\alpha_{23}$&1.48& 1.77& 2.09\\ 
\hline%\hline
$\alpha_{30}$&1.05& 1.11& 1.18&
$\alpha_{40}$&1.06& 1.10& 1.14\\ % \hline
$\alpha_{31}$&3.82& 4.99& 6.06&
$\alpha_{41}$&0& 0& 0\\ %\hline 
$\alpha_{32}$&11.0& 14.8& 18.1&
$\alpha_{42}$&1.07& 1.13& 1.21\\ %\hline
$\alpha_{33}$&1.15& 1.30& 1.48&
$\alpha_{43}$&1.09& 1.16& 1.24\\ 
\hline%\hline
$\alpha_{50}$&1.12& 1.19& 1.26&
$\alpha_{60}$&1.15& 1.25& 1.38\\  %\hline
$\alpha_{51}$&1.12& 1.19& 1.26&
$\alpha_{61}$&0&0&0\\ %\hline 
$\alpha_{52}$&1.10& 1.16& 1.24&
$\alpha_{62}$&0.92& 0.90& 0.88\\ %\hline
$\alpha_{53}$&1.18& 1.30& 1.42&
$\alpha_{63}$&1.24& 1.42& 1.64\\
\hline
$\alpha_{70}$&1.23&1.36&1.51&
$\alpha_{71}$&0.82&0.78&0.76\\ 
$\alpha_{72}$&1.25&1.42&1.62&
$\alpha_{73}$&1.35&1.57&1.82\\ 
\hline\hline
\end{tabular}
\label{tab1}
\caption{The factors $\alpha_{ab}$, introduced in Eq. (\ref{K-one-loop}), for typical numerical values $\Delta/M=0.2,\, 0.3,\, 0.4$. The analytic expressions for $\alpha_{ab}$ are given in the Appendix in Eq. (\ref{alphas-explicit}).}
\end{table}

A few comments on Eq. (\ref{K-one-loop}) are in order. The expansion of Eq. (\ref{def-expansion}) is rapidly converging with a suppression factor $\sqrt{t}/M_N \sim Q$, $m_\pi/M_N \sim Q$, or $\delta/M_N \sim Q^2$. The neglected terms of order $Q^6$ and higher lead to corrections of less than 1\% typically. In Table I we illustrate the importance of the $\Delta/M$ effect in the coefficients $\alpha_{ab}$ at three typical values for $\Delta/M = 0.2$, 0.3, 0.4. In most cases we find significant deviations from the limiting case $\alpha_{ab} \to 1$. Any attempt to recover this effect in terms of only a few moments appears futile. 

Particular attention should be paid to the terms in Eq. (\ref{K-one-loop}) proportional to $\alpha_{n0}$ and $\alpha_{0n}$. In the chiral domain they all violate the expectation from 
dimensional power-counting rules. We expect such terms to be cancelled by contributions from two-loop diagrams \cite{Long:2009wq,Lutz:2018cqo}. We therefore impose the renormalization condition
%\begin{eqnarray}
% \alpha_{n0} \to 0 \qquad {\rm  and } \qquad  \alpha_{0n} \to 0 \,.
%\end{eqnarray}
$ \alpha_{n0} \to 0$ and $\alpha_{0n} \to 0$, and also $\beta_{0n}\rightarrow 0$.

Where possible, we have compared our results for the loop contributions in Eq. \eqref{def-} with existing relativistic approaches \cite{Schindler:2006it, Yao:2017fym, Procura:2006gq}. We agree with Ref. \cite{Procura:2006gq}, which does not include isobar contributions, for bubble and tadpole diagrams, but we disagree with them for the triangle diagram with two internal nucleons. Our results agree with Eq. (21) in Ref. \cite{Schindler:2006it} if we make the replacement $16M_N^4I_{\pi NN}^{(PP)}(t) \rightarrow (16M_N^4-4tM_N^2)I_{\pi NN}^{(PP)}(t)$ and change the sign of the $g_R$-term.
In Ref. \cite{Yao:2017fym} results are given for the loop contributions to the axial charge, Eq. (A4), and the axial radius, Eq. (A5). By using Eqs. \eqref{ILpiRinbubbles} and \eqref{dILpiRdtinbubbles} from Appendix C we can rewrite triangle integrals in terms of bubble integrals. The full bubble contributions, denoted by $B_0(m_N^2,M_\pi^2,m_\Delta^2)$ and $B_0(m_\Delta^2,M_\pi^2,m_\Delta^2)$, can then be reproduced. The contributions proportional to the tadpole integral $A_0(M_\pi^2)$ differ by a minus sign from ours. We disagree with the normalization of the axial radius in Eq. (A5) by a factor 36.

\begin{figure}[h]
\includegraphics[width=150mm]{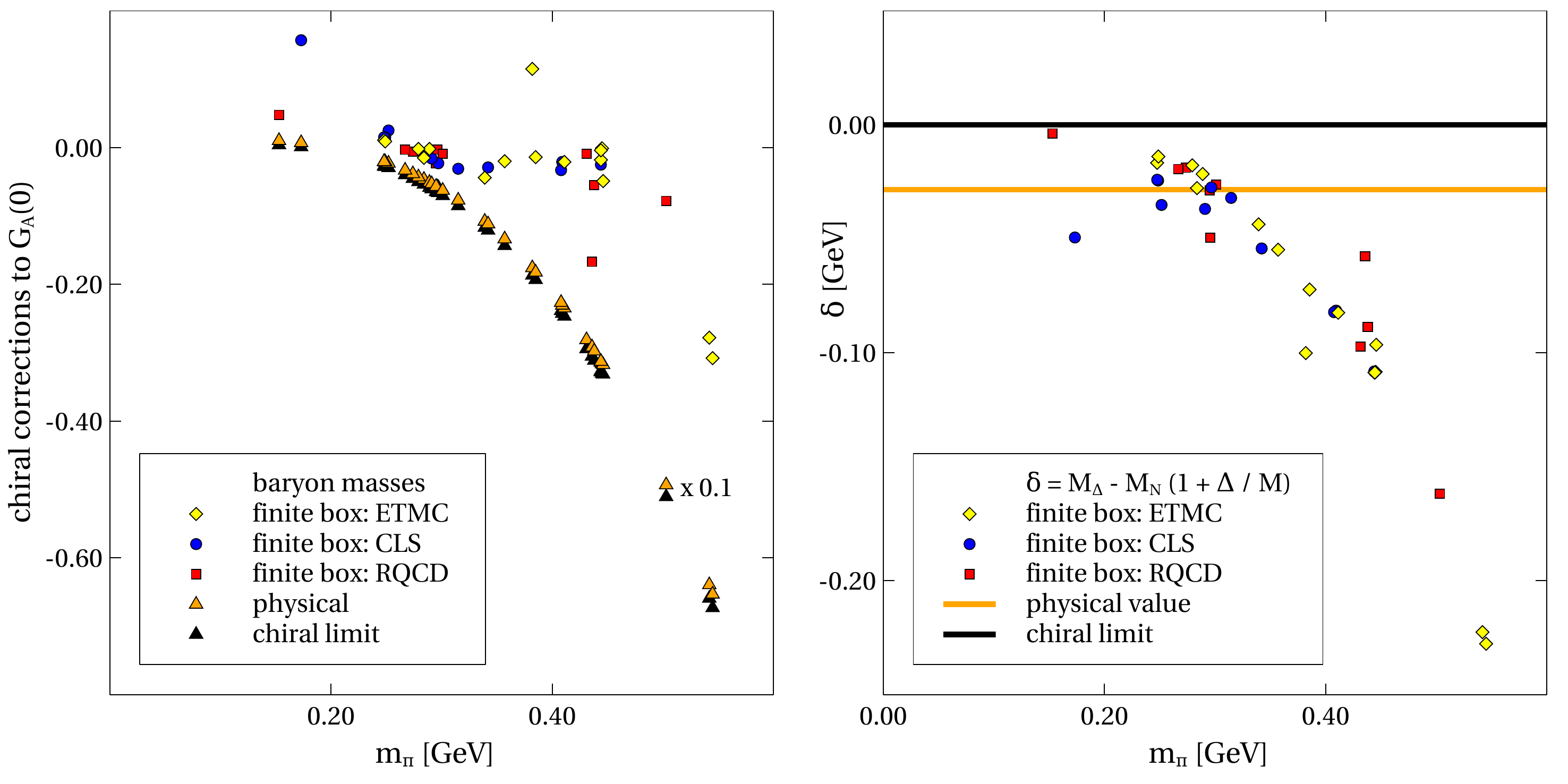}
\caption{A comparison between using chiral, physical, and on-shell baryon masses in the chiral correction terms to the axial charge. Since for the two cases with approximate baryon masses the chiral correction terms are quite large, their contributions are scaled down by a factor of 10. A strong effect is seen starting at pion mass $m_\pi \simeq 0.2$ GeV.}
\label{PlotgAQ4Contribution}
\end{figure} 

\section{Fit to QCD lattice data}
\label{sec:LQCD}
As a first application of our results we consider the set of QCD lattice data on the nucleon and isobar mass (when available) and the nucleon axial-vector form factor. We use the evolutionary fit algorithm GENEVA 1.9.0-GSI \cite{Geneva}. We take into account the most recent two-flavor ensembles of the ETMC \cite{Alexandrou:2008tn,Alexandrou:2010hf}, which provides also data for the isobar mass, the CLS \cite{Capitani:2017qpc,Capitani:2015sba}, and the RQCD \cite{Bali:2014nma}  QCD lattice collaborations.

\begin{table}
\centering%\scriptsize
\renewcommand{\arraystretch}{1.0}
\begin{tabular}{|c | c |c c|c c|}\hline\hline
group&scale & this work& lattice group
\\ \hline
ETMC&$a_{\beta=3.8}\,[\text{fm}]$&$0.1075(_{-0.0005}^{+0.0010})$&
$0.0995(7)$ \cite{Alexandrou:2008tn}
\\ \cline{2-4}
&$a_{\beta=3.9}\,[\text{fm}]$&$0.0932(_{-0.0006}^{+0.0004})$& $0.089(5)$ \cite{Alexandrou:2010hf}
\\ \cline{2-4}
&$a_{\beta=4.05}\,[\text{fm}]$&$0.0731(_{-0.0002}^{+0.0004})$&
$0.070(4)$ \cite{Alexandrou:2010hf}
\\ \cline{2-4}
&$a_{\beta=4.2}\,[\text{fm}]$&$0.0587(_{-0.0007}^{+0.0003})$&
$0.056(4)$ \cite{Alexandrou:2010hf}
\\ \hline
CLS&$a_{\beta=5.2}\,[\text{fm}]$&$0.0841(_{-0.0006}^{+0.0007})$&
0.079  \cite{Capitani:2017qpc}
\\ \cline{2-4}
&$a_{\beta=5.3}\,[\text{fm}]$&$0.0703(_{-0.0006}^{+0.0006})$ &
0.063 \cite{Capitani:2017qpc}
\\ \cline{2-4}
&$a_{\beta=5.5}\,[\text{fm}]$&$0.0525(_{-0.0004}^{+0.0005})$&
0.050 \cite{Capitani:2017qpc}
\\ \hline
RQCD&$a_{\beta=5.2}\,[\text{fm}]$&$0.0830(_{-0.0009}^{+0.0007})$& 
0.081 \cite{Bali:2014nma}
\\ \cline{2-4}
&$a_{\beta=5.29}\,[\text{fm}]$&$0.0700(_{-0.0003}^{+0.0005})$&
0.071 \cite{Bali:2014nma}
\\ \cline{2-4}
&$a_{\beta=5.4}\,[\text{fm}]$&$0.0588(_{-0.0006}^{+0.0002})$&
0.060 \cite{Bali:2014nma}
\\ \hline\hline
\end{tabular}
\caption{Lattice scales as determined in our fit.}
   \label{LatticeaFits}
\end{table}

Our fit strategy is the following. We fit our expressions for the nucleon and isobar masses, $M_N$ and $M_\Delta$ in Eq. (\ref{def-sigma}) (with finite-volume effects included following Ref. \cite{Lutz:2014oxa}), and for the nucleon axial-vector form factor, $G_A(t)$ in Eq. (\ref{def-}) (without explicit finite-volume effects), to the lattice data. As explained, we use on-shell baryon masses in the loops, in order to keep the analytical structures of the integrals. We include the $K$ factors of Eq. (\ref{K-one-loop}), but not Eq. (\ref{K-higher-order}). This means that higher-order terms $\sim Q^4$ are not taken into account. We checked that these higher-order contributions are small for low-enough pion masses, $m_\pi < 0.55$ GeV. These convergence properties are dramatically improved by the use of on-shell masses as illustrated in Figure \ref{PlotgAQ4Contribution}. There we evaluate the chiral correction terms of the axial charge on all considered lattice ensembles by means of Eqs. (\ref{def-}) and (\ref{K-one-loop}). 
The results depend on the LECs, the pion mass of the ensemble, and the baryon masses. In the Figure three different choices for the baryon masses are illustrated. While within our on-shell scheme we find reasonably-sized chiral correction terms up to pion masses of a 500 MeV, this is not the case for the conventional case of using chiral-limit baryon masses in the loop functions or physical values as was used in Ref. \cite{Yao:2017fym}. The source of the large differences  is due to distinct values for $\delta$ in the coefficients $K$ (see the right-hand panel of the Figure), but also the use of the finite-box baryon masses in the scalar loop integrals as they are predicted on the various lattice ensembles by our global fit. The scatter in our results reflects the decisive role of finite-volume effects on the baryon masses. 
 
It is important to have accurate values for the QCD lattice scales available on all considered lattice ensembles. There are different scale-setting schemes used by the 
lattice collaborations, which differ by the size of the discretization effects. We follow here the path of ETMC \cite{Alexandrou:2008tn}, which suggests to 
set the scale by the requirement to recover the isospin-averaged mass of the nucleon at the physical point. With this construction discretization effects are minimized in the baryon masses. Since this relies on a particular chiral extrapolation scheme, we consider the various lattice scales as free parameters in our global fit. Such a strategy was successfully used in various global fits to lattice data  \cite{Lutz:2014oxa,Lutz:2018cqo,Guo:2019nyp}. Our results for all lattice scales are shown in Table \ref{LatticeaFits}. The scales given by the ETMC and CLS collaborations differ significantly from our values. There is, however, a clear trend that the smaller the lattice scale, the closer our fitted scales get to the ones given by the lattice collaborations. The RQCD scales, on the other hand, can be reproduced quite accurately. 
Since the lattice set up of the CLS and RQCD groups coincide, one would expect 
identical lattice scales on the $\beta =5.2$ ensembles in Table \ref{LatticeaFits}. Within uncertainties this is the case for our results.

We include in the fit the lattice points used up to pion mass $m_\pi=$ 0.55 GeV for the nucleon and isobar masses and the nucleon axial-vector form factor. For the form factor we include the data points up to momentum transfer $t= 0.36$ GeV$^2$ and for lattice sizes with $m_\pi L \geq 4.0$, where we expect good convergence properties and (explicit) finite-volume effects to be small. In the plots all data points are shown, but we identify and reject two outliers, {\it viz.} the masses connected to the highest pion masses of ETMC ($a=0.1075 \,\text{fm}$, $m_\pi=0.542$ GeV) and the highest nucleon mass of RQCD ($a=0.0700\, \text{fm}$, $m_\pi=0.436$ GeV). For both points there is no $G_A$ data to be fitted, since there is none available for ETMC ($a=0.1075$ fm) and $m_\pi \, L < 4.0$ for RQCD ($a=0.0700\, \text{fm}$, $m_\pi=0.436$ GeV).

\begin{table}
\centering%\scriptsize
\renewcommand{\arraystretch}{1.0}
\begin{tabular}{c c |c c| c c}\hline\hline
LEC&Fit result&LEC&Fit result&LEC&Fit result
\\ \hline
$f \,[\text{MeV}]$&$87.19(_{-0.19}^{+0.24})$&
$b^*_\chi \,[\text{GeV}^{-1}]$&$-0.839(_{-0.011}^{+0.006})$&
$g_S\,[\text{GeV}^{-1}]$&$0.433(_{-0.057}^{+0.081})$
\\ \hline
$M \,[\text{MeV}]$&$884.80(_{-0.72}^{+0.36})$&
$d^*_\chi \,[\text{GeV}^{-1}]$&$-0.556(_{-0.006}^{+0.011})$&
$g_V\,[\text{GeV}^{-2}]$&$-1.133(_{-0.434}^{+0.117})$
\\ \hline
$M+\Delta \,[\text{MeV}]$&$1187.09(_{-0.30}^{+0.58})$&
$c_\chi \,[\text{GeV}^{-3}]$&$2.308(_{-0.040}^{+0.067})$&
$g_T\,[\text{GeV}^{-1}]$&$1.554(_{-0.054}^{+0.097})$
\\ \hline
$g_A$&$1.1933(_{-0.0032}^{+0.0036})$&
$e_\chi \,[\text{GeV}^{-3}]$&$1.461(_{-0.027}^{+0.022})$&
$g_R \,[\text{GeV}^{-2}]$&$0.925(_{-0.024}^{+0.032})$
\\ \hline
$f_S$&$1.9409(_{-0.0145}^{+0.0097})$&
$\bar{l}_3$&$3.255(_{-0.089}^{+0.043})$&
$h_S\,[\text{GeV}^{-1}]$&$-0.246(_{-0.065}^{+0.073})$
\\ \hline
$h^*_A$&$-0.9057(_{-0.0563}^{+0.0964})$&
$g_\chi \,[\text{GeV}^{-2}]$&$-3.597(_{-0.083}^{+0.034})$&
$h_V\,[\text{GeV}^{-2}]$&$-1.357(_{-0.149}^{+0.152})$
\\ \hline
&&&&$f_A^{(3)}+f_A^{(4)}/5 \,[\text{GeV}^{-1}]$&$-4.139(_{-0.033}^{+0.064})$
\\ \hline\hline
\end{tabular}
\caption{Low-energy constants as determined in our fit. The $*$ parameters are not fitted to the lattice data. While $b_\chi$ and $d_\chi$ are adjusted to the isospin-averaged masses of the nucleon and the isobar at the physical point, the value of $h_A = 9\,g_A-6\,f_S$ is implied by its large-$N_c $ sum rule.}
   \label{LECsFits}
\end{table}

The fit minimizes the least-squares differences $\chi^2$ of our expressions with respect to the lattice data points. In this $\chi^2$ determination, all available lattice points that meet our requirements (see previous section) contribute with equal weight. The $\chi^2$ per lattice point reached is $\chi^2_{\rm min}/N_{\rm data}=1.04$. With $99$ used lattice data points and 26 degrees of freedom 19-3 LECs and 10 lattice scales), we reach for the total $\chi^2$ per degree of freedom
\begin{align}
\chi^2_{\rm min}/N_{\rm df} =103.0/(99-26) = 1.40 \, ,  
\end{align}
which qualifies as a good description of the available lattice data. We give asymmetric error bars. They are based on a one standard deviation ($\sigma$) change for the value of $\chi^2_{\rm min}$ ({\it i.e.} an increase by 1). We determined the region for the LECs meeting this range, from which follow the errors for the LECs.

\begin{table}[t]
\centering%\scriptsize
\renewcommand{\arraystretch}{1.0}
\begin{tabular}{c| c ||c |c}\hline\hline
LEC& Tree-level matching \cite{Guo:2019nyp} &LEC & Tree-level matching \cite{Guo:2019nyp} \\  \hline 
$f_S$          &1.50     &$h_A$          &2.07                    \\ \hline
$b_{\chi}$     &$-$0.85  &$d_{\chi}$     &$-$0.64                  \\ \hline
$c_{\chi}$     &0.20     &$e_{\chi}$     &$-$0.18                  \\ \hline
$g_V$          &0.80     &$h_V$          &2.05                     \\ \hline
$g_S$          &$-$3.04  &$h_S$          &$-$3.22                  \\ \hline
\end{tabular}
\caption{A set of LECs from a flavor-SU(3) analysis.} 
\label{CompareSU3}
\end{table} 

The LECs determined in the fit are collected in Table \ref{LECsFits}. Of the large-$N_c$ relations in Eq. (\ref{LargeNcSU2}) we use for now only the relation that eliminates $h_A$. In addition, we set  $\zeta_N=\zeta_\Delta=0$, because in the axial-form factor their effect can be renormalized into the value of $g_\chi$. We note that in our current work the impact of the large-$N_c$ relations is quite limited, because the data set that we consider does not constrain most of the isobar parameters $h^{(n)}_S$ and $h^{(n)}_V$. Given our fit results, Eq. (\ref{LargeNcSU2})  can be used to derive an estimate of the LECs that our data set is not directly sensitive to. We find reasonable values for $f$, $M$, $M+\Delta$, and $g_A$  \cite{Aoki:2019cca}. It is noteworthy that $f_S=1.94$ takes a large value, leading to a negative value for $h_A$, which was also reported in Ref. \cite{Yao:2016vbz} (denoted as $g_1$). The middle column of Table \ref{LECsFits} lists chiral-symmetry-breaking LECs. It shows an expected value for $\bar{l}_3$ \cite{Bernard:2007zu,Aoki:2019cca}. The leading-order large-$N_c$ relations $b_\chi=d_\chi$ and $c_\chi=e_\chi$ are approximately fulfilled.  The value of $g_\chi$ is poorly known in literature. The right column shows the chiral-symmetry-conserving LECs of higher order. We find values for $g_S$ and $g_V$ that disagree significantly from previous SU(2) works like Refs. \cite{Bernard:2007zu,Gasparyan:2010xz}. However, in most papers the constants $g_S$ and $g_V$ are determined in a theory without isobars. How to translate this to our case with isobars is not clear (see page 10 of Ref. \cite{Bernard:2009mw}). 
The other LECs of this column have not been determined reliably in the literature so far. In general, our determined LECs are in the expected range and they appear to be reasonably small. The one-sigma error bars in the LECs are rather small in general, with a notable exception of $g_V$, for which its size does not seem to depend very strongly 
on the baryon masses and form factor.

It is illuminating to also compare our set of LEC with results from flavor-SU(3) analyses. Here we focus on the most recent work \cite{Guo:2019nyp}, which achieved a global fit to the baryon octet and decuplet masses as provided by various lattice groups. In Table \ref{CompareSU3} we collect values that are obtained by a tree-level matching of the flavor-SU(2) with the flavor-SU(3) chiral Lagrangian. A comparison of Table \ref{CompareSU3} with Table \ref{LECsFits} reveals an interesting pattern. While the LECs $b_\chi$ and $d_\chi$ at order $Q^2$ are quite compatible, the higher-order terms $c_\chi$ and $e_\chi$ are quite distinct, which is not necessarily surprising or problematic. Most striking is the opposite sign
in $h_A$, which we interpret as a signal for the importance of strangeness loop effects. This is in line with a striking prediction of the flavor-SU(3) analyses \cite{Lutz:2018cqo,Guo:2019nyp}, which obtained a strangeness sigma term of the isobar  $\sigma_{s\Delta} \simeq -270$ MeV, significantly larger in magnitude  than its corresponding value for the nucleon with $\sigma_{sN} \simeq 45$ MeV (here we provide unpublished results from Ref. \cite{Guo:2019nyp}).

\begin{figure}[t]
\includegraphics[width=120mm]{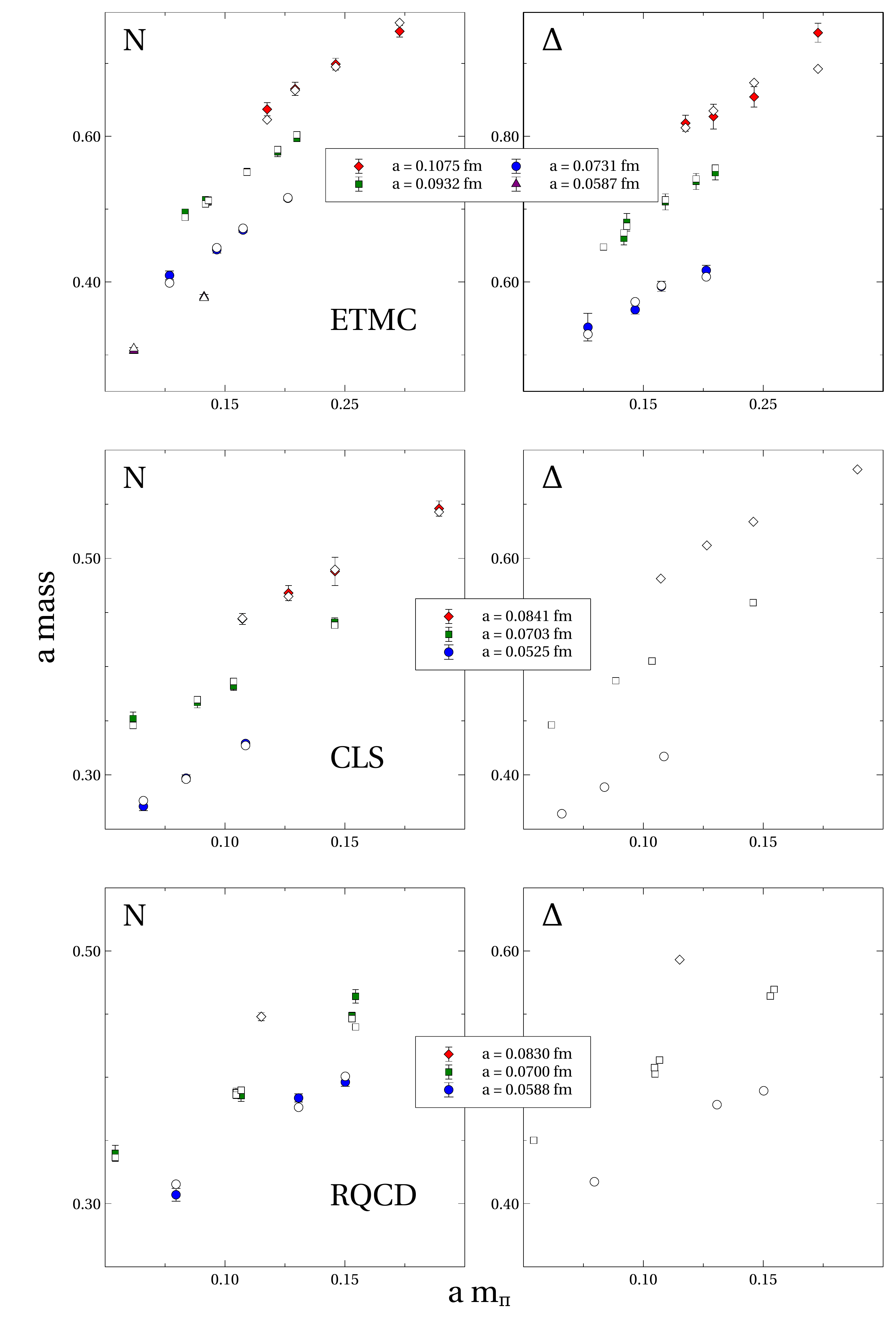}
\caption{Comparison between our (white) theory results for the nucleon and isobar masses and (colorful) results of different lattice collaborations. We give our fitted lattice scales as identification. The two outliers are the highest red point of ETMC and the highest green point of RQCD.}
\label{PlotMasses}
\end{figure}

The masses of the nucleon and the isobar are reproduced excellently (see Figure \ref{PlotMasses}). The colored points represent the lattice data of the different ensembles and the white points are our theory predictions for the corresponding lattice volume. Most theory error bars are too small to be shown. The collaborations CLS and RQCD do not give results for the isobar mass $M_\Delta$. In these cases, the white points are to be seen as predictions.

\begin{figure}[h]
\includegraphics[width=140mm]{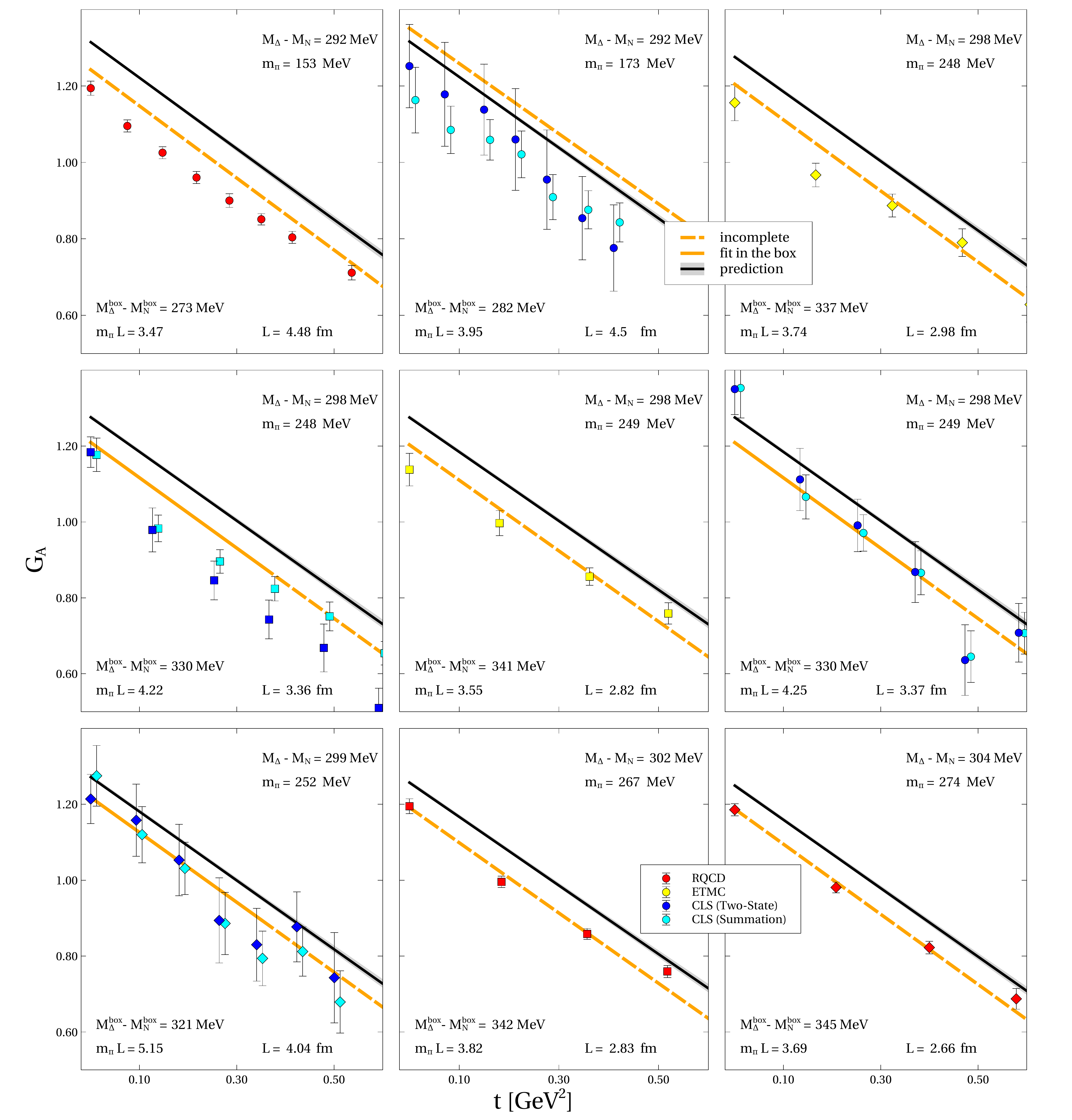}
\caption{The axial-vector form factor $G_A(t)$ of the nucleon for different pion masses. The colored lattice points are to be compared to our finite-box results (the straight orange lines). The dashed lines visualize the lattice points that are not fitted, as explained in the text. The black lines represent our results in the infinite-volume limit.}
\label{PlotGA1}
\end{figure}

\begin{figure}[h]
\includegraphics[width=140mm]{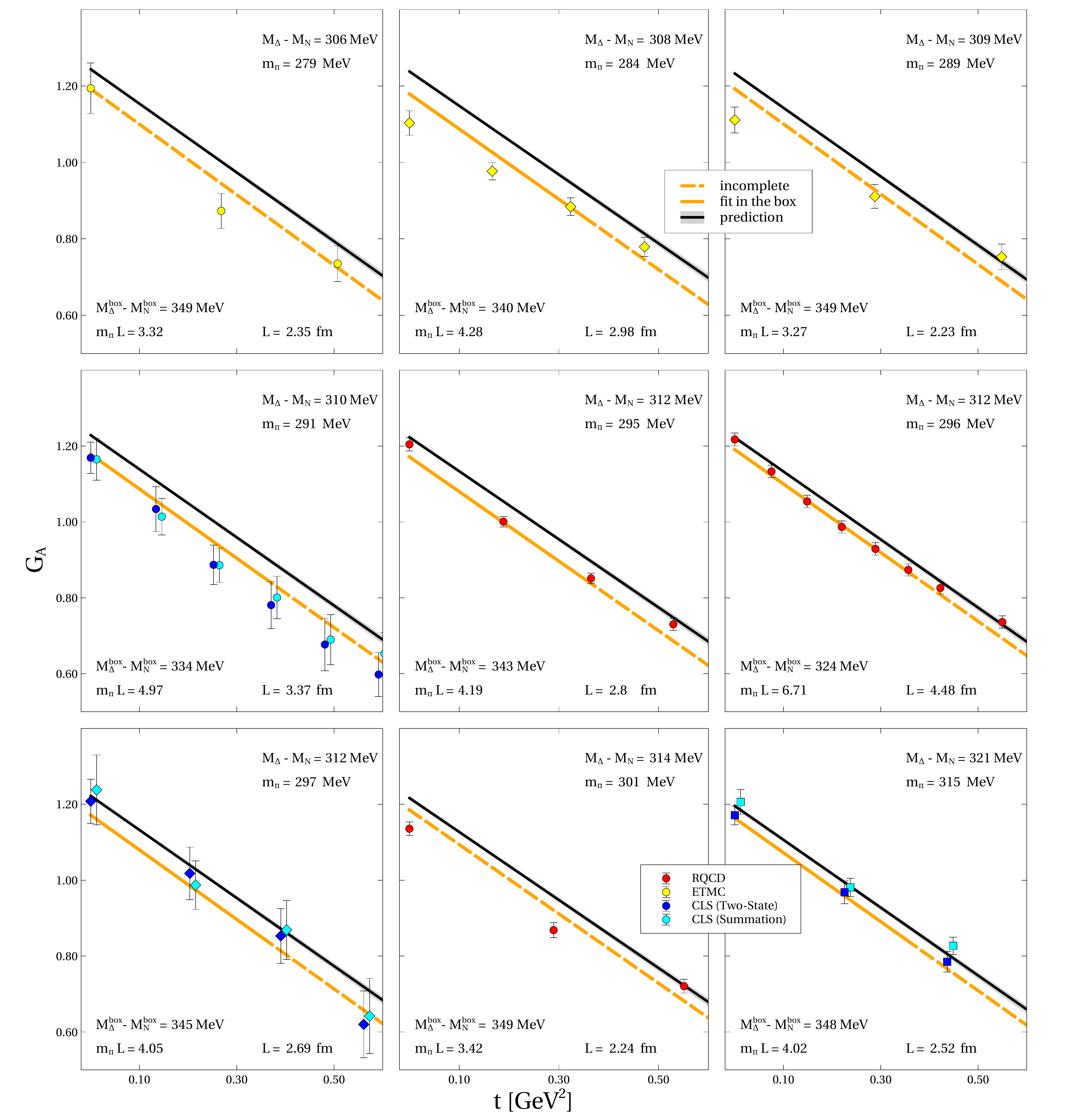}
\caption{The axial-vector form factor $G_A(t)$ of the nucleon for different pion masses; see the caption of Figure \ref{PlotGA1}.}
\label{PlotGA2}
\end{figure}

\begin{figure}[h]
\includegraphics[width=140mm]{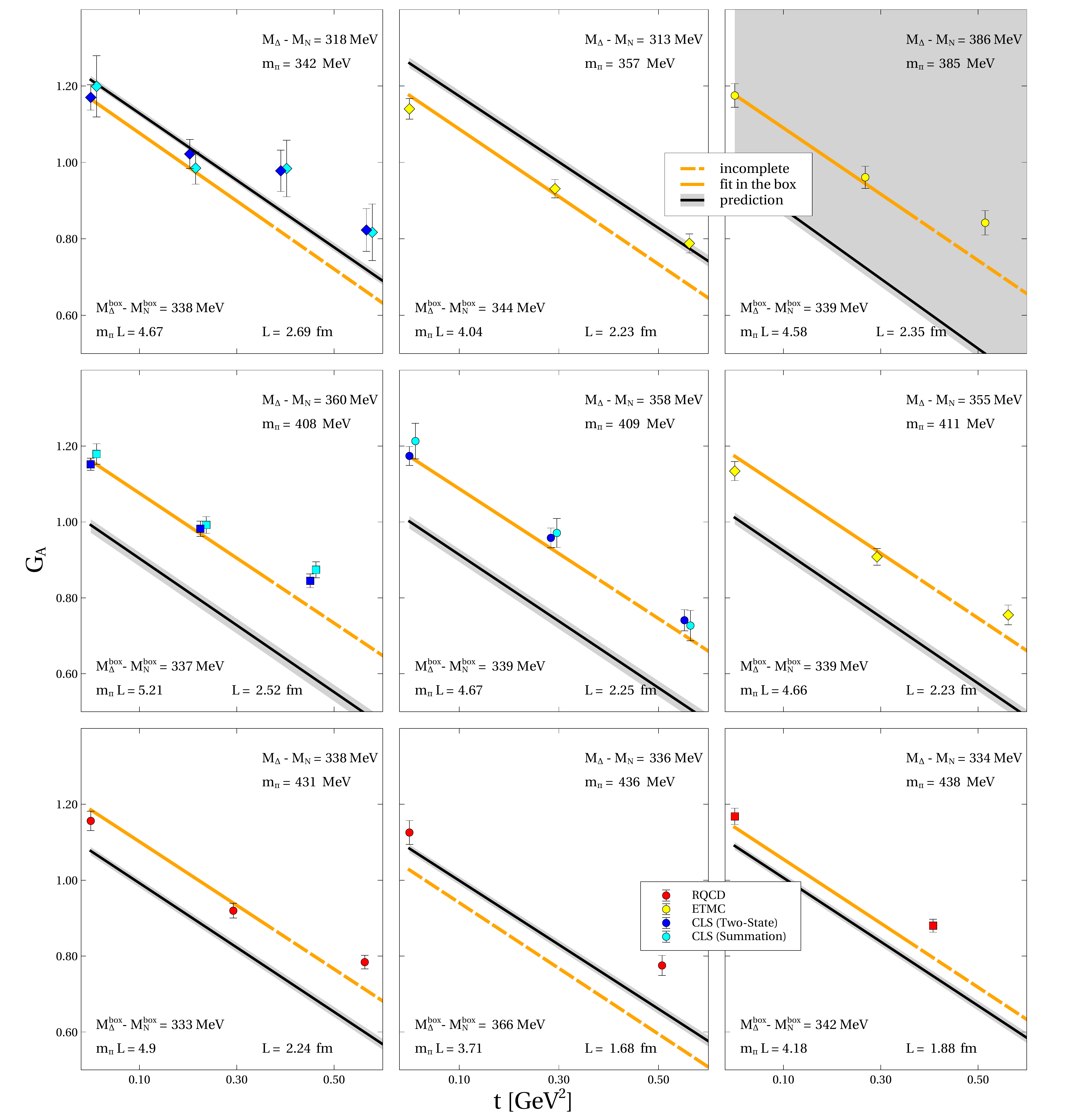}
\caption{The axial-vector form factor $G_A(t)$ of the nucleon for different pion masses; see the caption of Figure \ref{PlotGA1}.}
\label{PlotGA3}
\end{figure}

\begin{figure}[h]
\includegraphics[width=140mm]{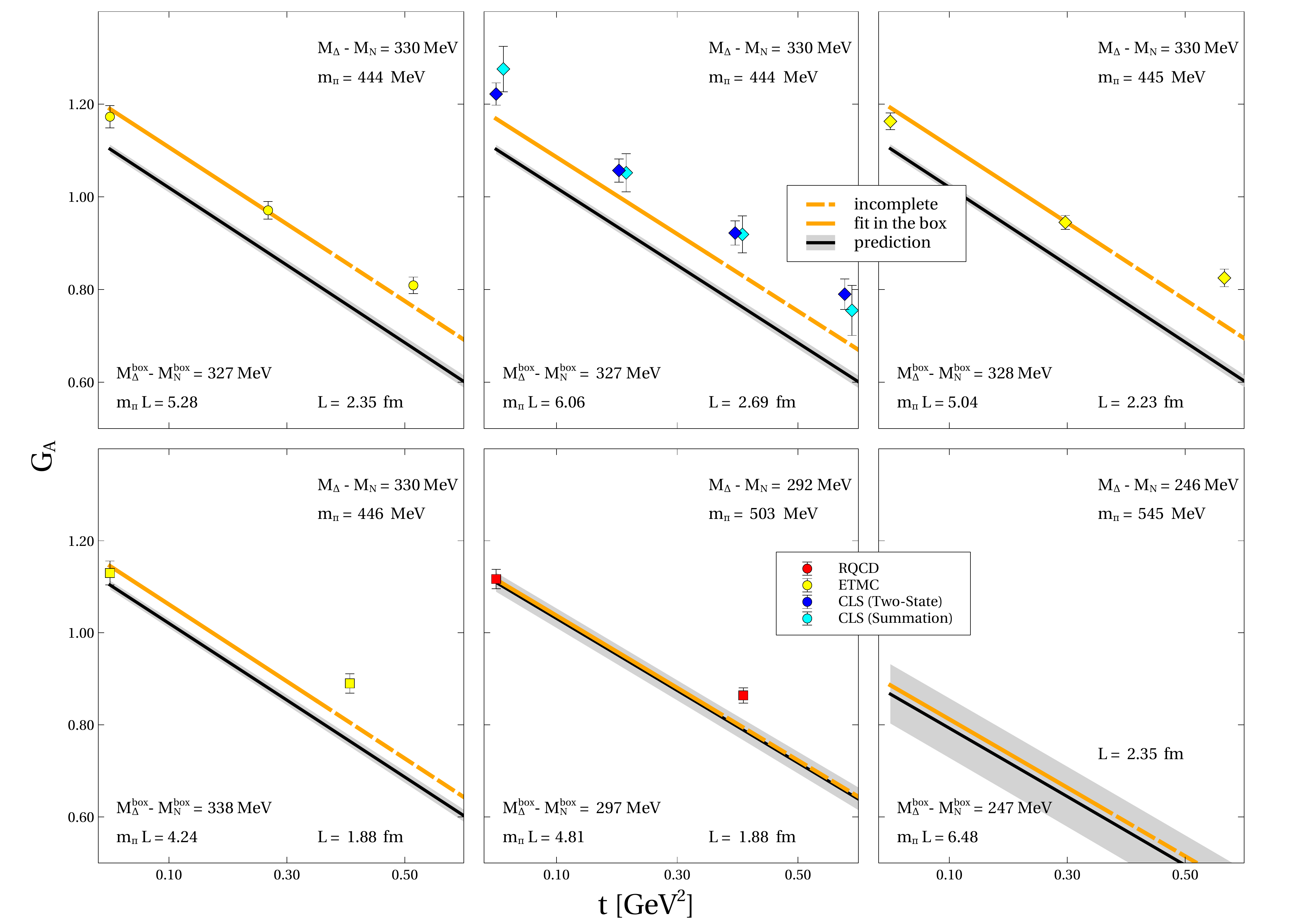}
\caption{The axial-vector form factor $G_A(t)$ of the nucleon for different pion masses; see the caption of Figure \ref{PlotGA1}.}
\label{PlotGA4}
\end{figure}

Figures \ref{PlotGA1}-\ref{PlotGA4} show the axial-vector form factor of the nucleon $G_A(t)$ for each lattice point. The order is given by increasing pion mass. The color of the points determines the corresponding lattice group, whereas the used symbol indicates the size of the lattice scale $a$ (diamond-circle-square, from largest to smallest). We present our theoretical results in the infinite-volume limit (black line with gray error band) and extrapolated to the box size of the corresponding lattice point (orange lines).
Finite-volume effects of the form factor originate only from finite-volume effects of the masses, which enter the expressions for the loop contributions. Explicit finite-volume effects, originating directly from the loop integrals of the form factors, are not taken into account.
In order to visualize the lattice points which do not enter the fit  due to the restrictions described above ($m_\pi \, L < 4.0$ and $t>0.36\,\text{GeV}^2$), we used dashed orange lines as compared to the fitted data points which are indicated by solid orange lines.
For convenience we also show the isobar-nucleon mass gap in the infinite volume and the pion mass in the top-right corner and the mass gap in the box and the lattice size $m_\pi \, L$ and $L$ in the bottom of each plot.
The plots of the data points of the CLS collaboration include the fitted ``Two-State Method'' (dark blue) and the light blue ``Summation'' method, which does not enter the fit.
We find a very good description of the lattice points with the corresponding solid orange theory results. 
%\textbf{\textit{(Note that some lattice data seems contradictive, e.g. the CLS plots with $m_\pi = 0.248/0.249$ GeV at lattice size $L=3.36/3.37$ fm in figure \ref{PlotGA1}!)}}

\begin{figure}[h]
\includegraphics[width=150mm]{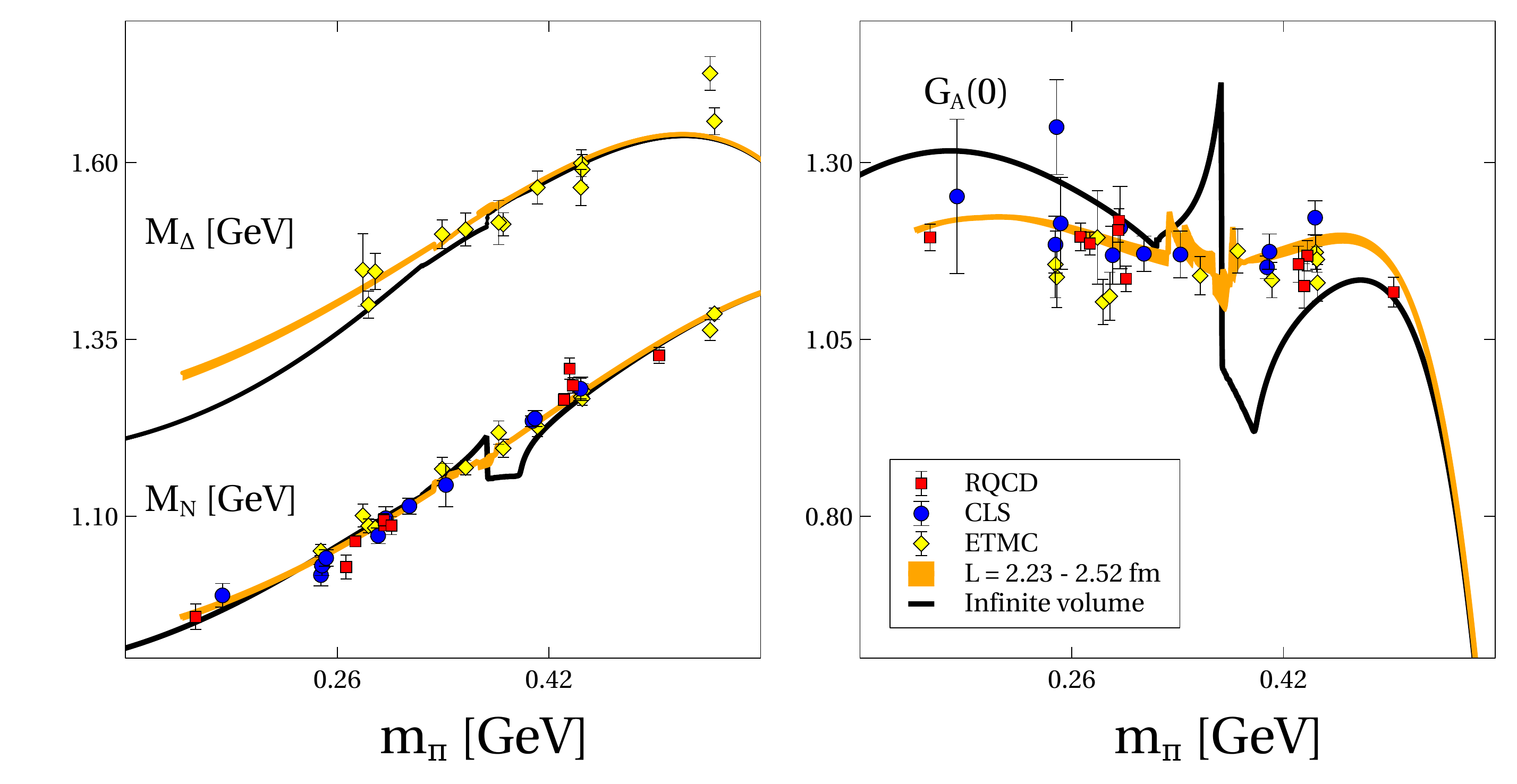}
\caption{Nucleon and isobar masses and the axial-vector form factor of the nucleon as function of the pion mass compared to the QCD lattice data. The orange lines are our results for the indicated range of lattice sizes, the black lines are our predictions in the infinite-volume limit.}
\label{PlotJump1}
\end{figure}

\begin{figure}[h]
\includegraphics[width=150mm]{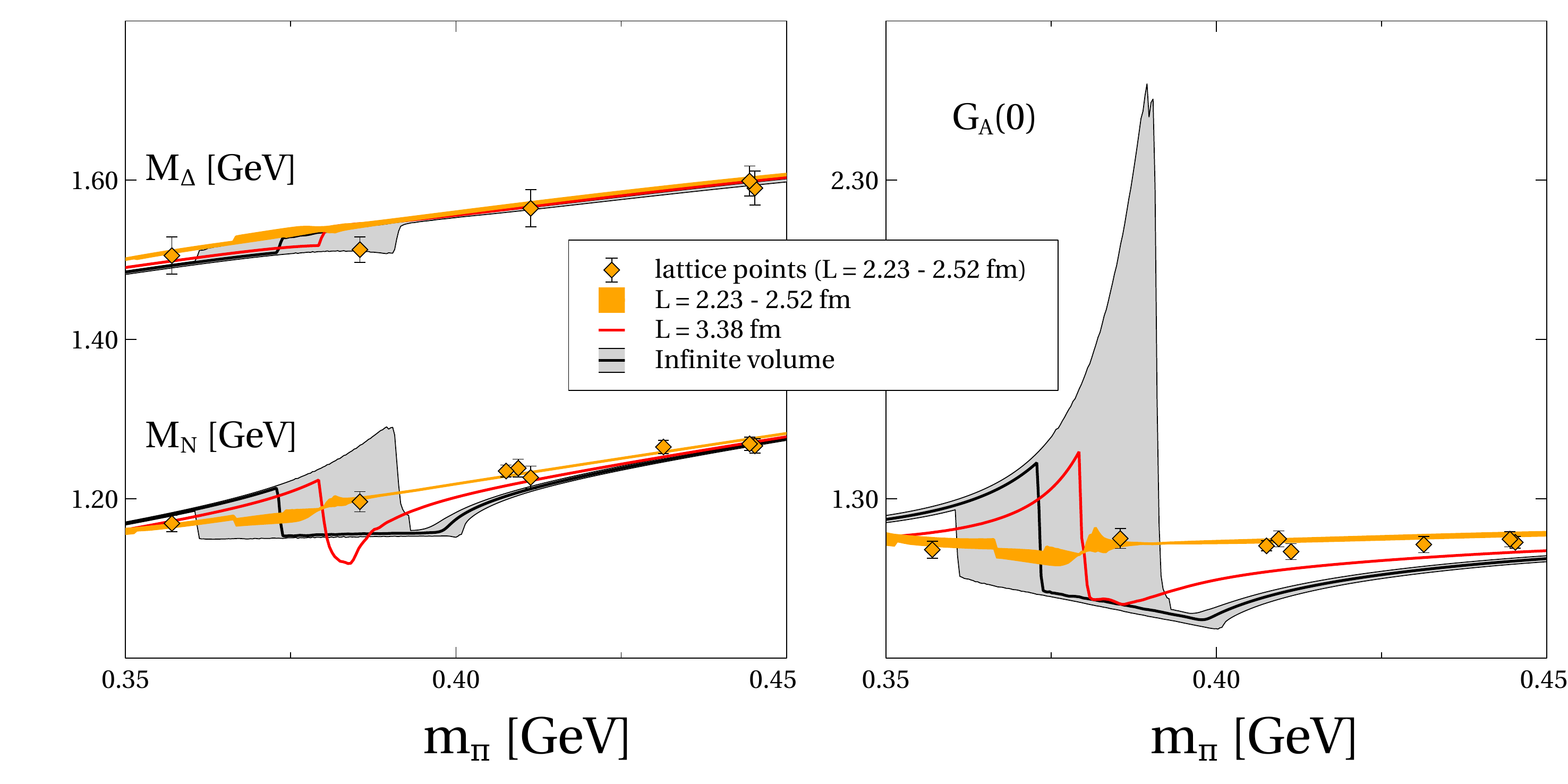}
\caption{Nucleon and isobar masses and the axial-vector form factor of the nucleon as function of the pion mass compared to the QCD lattice data, zoomed in to the region where our results show the non-analytic behavior. The orange lines are our results for the indicated range of lattice sizes, the red lines are our results for the indicated larger lattice size, and the black lines are our predictions in the infinite-volume limit.}
\label{PlotJump2}
\end{figure}

A large error band of the infinite volume prediction appears in the upper-right corner of Figure \ref{PlotGA3} at pion mass $m_\pi=0.385$ GeV.
This plot is particularly interesting, because $M_\Delta^{\text{box}}-M_N^{\text{box}}<m_\pi<M_\Delta-M_N$. This means that the isobar is unstable in our theory, but stable in the lattice simulation, implying a different analytical behavior.
In order to scrutinize this in more detail, we provide Figures \ref{PlotJump1} and \ref{PlotJump2}, which show the full range of our results for the nucleon and isobar masses and the form factor at the physical point $g_A\equiv G_A(t=0)$. 

In general, the region around $m_\pi=375\,\text{MeV}$ seems is very interesting, because, when varying $M_N$, $M_\Delta$ and $g_A$ in terms of the pion mass, we find a clear jump. This has been observed before \cite{Semke:2006hd, Guo:2019nyp} in SU(3) and is nicely displayed in Figures \ref{PlotJump1} and \ref{PlotJump2}.
The full treated range is shown in Figure \ref{PlotJump1}, where we confront the lattice data with our theory in finite (orange, $L\in [2.23, 2.52]\, \text{fm}$) and infinite (black) volume. We find a large difference between the finite volume and the infinite volume, especially for the axial charge $G_A(0)$. We keep in mind that the lattice points are not expected to match with any of the given lines if their volume $L\not\in [2.23, 2.52]\, \text{fm}$. 

We see a clear jump around $m_\pi = 375 \,\text{MeV}$ in all three observables. Figure \ref{PlotJump2} allows a closer look into that. We show  our results in infinite volume (black with gray error band), large lattice volume (red), and small lattice volume (orange). Additionally we show lattice results, which are unfortunately only available for small lattice sizes. We find a rather smooth curve for small box size, but there are clear jumps for larger boxes, but especially in the infinite volume. In the infinite-volume limit we determine the position of the jump as $m_\pi=373.49(_{-12.30}^{+17.79}) \, \text{MeV}$ and its height for the nucleon/isobar mass at $60.13(_{-23.17}^{+65.48})\, \text{MeV}/\;18.84(_{-4.38}^{+17.74})\, \text{MeV}$. This means that we report 3-4 $\sigma$ evidence for the existence of this jump. Our plot shows that lattice size  $L\geq 3.38$ fm should be sensible to this jump, especially when determining the axial charge $G_A(0)$. We therefore encourage the lattice community to investigate this suggested behavior, which is a direct consequence of our scheme, namely the use of on-shell masses in the loop contributions and the self-consistent determination of the baryon masses. Conventional chiral perturbation theory approaches, which use expanded masses in the loops, will not see this phenomenon, because approximated solutions of Eq. (\ref{def-sigma}) are always linear and do not allow for non-linear, self-consistent equations that result from the use of on-shell masses in the loops.

The observables that follow from our fit are summarized in Table \ref{ObservablesFits}. 
The experimentally well-known parameter $G_A(0)=1.2732(23)$ \cite{Tanabashi:2018oca} is reproduced quite accurately. 
We have to keep in mind that this flavor-SU(2) approach has limited validity, because strangeness loops may be relevant in specific observables. 
Our result for the nucleon axial radius is $\langle r_A^2\rangle=0.16656(_{-0.00348}^{+0.00322})\text{fm}^2$, which is pretty small, compared to $\langle r_A^2\rangle =0.263(38)\text{fm}^2$ \cite{Yao:2017fym} and the experimental data, $\langle r_A^2\rangle =0.46(22)\text{fm}^2$ \cite{Meyer:2016oeg} and $\langle r_A^2\rangle =0.46(24)\text{fm}^2$ \cite{Hill:2017wgb}.
The available lattice results are $\langle r_A^2\rangle =0.266(17)(7)\text{fm}^2$ \cite{Alexandrou:2017hac}, $\langle r_A^2\rangle =0.360(36)(^{+80}_{-88})\text{fm}^2$ \cite{Capitani:2017qpc}, and $\langle r_A^2\rangle =0.213(6)(13)(3)\text{fm}^2$ \cite{Green:2017ke}.

The sigma terms are defined by
\begin{align}
 \sigma_j=m\, \frac{\partial}{\partial \,m}\, M_j \ .
\end{align}
We determine $\sigma_N=49.305(_{-0.124}^{+0.409})$ MeV. 
A recent study of ETMC \cite{Alexandrou:2019brg} suggests that $\sigma_N=41.6(3.8)$ MeV. The RQCD collaboration \cite{Bali:2016lvx} finds a lower value, $\sigma_N=35(5)$ MeV. We find good agreement with the sigma terms from the extensive flavor-SU(3) mass fits \cite{Lutz:2018cqo,Guo:2019nyp} $\sigma_N\simeq 48$-$49$ MeV, $\sigma_\Delta\simeq 37$-$42$ MeV. We recall also the previous analysis \cite{Procura:2006bj,Alvarez-Ruso:2013fza} that obtained $\sigma_{\pi N}$ = 41(5)(4) MeV based on a flavor-SU(2) extrapolation of an older set of lattice data for the nucleon mass \cite{Bali:2012qs,Alexandrou:2010hf,Engel:2010my,Capitani:2012gj}. A comparison with the empirical value $\sigma_{\pi N} = 58(5)$ MeV from Ref. \cite{RuizdeElvira:2017stg} is of limited use for us, because it is not clear how important strangeness effects are.

 \begin{table}
\centering%\scriptsize
\renewcommand{\arraystretch}{1.0}
\begin{tabular}{c |c }\hline\hline
Observable&Fit results
\\ \hline
$G_A(0)$&$1.3094(_{-0.0042}^{+0.0044})$
\\ \hline
$\langle r_A^2\rangle\,[\text{fm}^2]$&$0.1666(_{-0.0035}^{+0.0032})$
\\ \hline
$\sigma_N\,[\text{MeV}]$&$49.31(_{-0.12}^{+0.41})$
\\ \hline
$\sigma_\Delta\,[\text{MeV}]$&$45.19(_{-0.44}^{+0.18})$
\\ \hline
jump position [MeV]&$373.5(_{-12.3}^{+17.8})$
\\ \hline
jump height nucleon [MeV]&$60.1(_{-23.2}^{+65.5})$
\\ \hline
jump height isobar [MeV]&$18.8(_{-4.4}^{+17.7})$
\\ \hline\hline
$\chi^2_{\rm min}/N_{\rm data}$ &1.04
\\ \hline\hline
$\chi^2_{\rm min}/N_{\rm df}$& 1.40
\\ \hline\hline
\end{tabular}
\caption{Observables as determined in our fit.}
   \label{ObservablesFits}
\end{table}

\clearpage
\section{Conclusion and outlook}

In this work we studied QCD with up and down quarks. From an effective field theory point of view baryonic systems are of particular interest, since here the intricate interplay of the prominent low-energy scales, the pion mass, $m_\pi$, and the isobar-nucleon mass difference, $\Delta $, can be scrutinized in the absence of additional complications from the strange quark. 

We considered QCD lattice data with two dynamical quark fields on the nucleon mass, the isobar mass, as well as the axial-vector form factor of the nucleon. A global fit to such data was performed successfully, as application of the two-flavor chiral Lagrangian. 
Accurate results are obtained for pion masses up to 500 MeV with a $\chi_{\rm min}^2/N_{\rm{df} } \simeq 1.40$. We illustrated that this became possible only because we applied 
an effective field theory framework that insists on the use of on-shell masses in the 
one-loop chiral correction terms. As an unavoidable consequence of such an approach we 
predict a discontinuous quark-mass dependence of the baryon masses and form factors at a 
pion mass $m_\pi =373(^{+18}_{-12})$ MeV, when evaluated in the infinite-box limit.
It was illustrated that at box sizes of current QCD lattice ensembles such a phase transition is not visible. However, it should be easily detectable if ensembles on 64$^3$ lattices in that pion-mass region are generated. We suggest to measure the nucleon {\it and} the isobar finite-box masses on such ensembles, since both enter the chiral dynamics of that system decisively. 

Further work is required to consolidate our results. So far we only considered 
volume effects that arise from an evaluation of the finite-box baryon masses.
Such effects are instrumental to recover the axial-form factor results on the various QCD lattice ensembles, and it remains for us to implement explicit finite-volume 
effects in our form-factor computation. A generalization of our framework to baryon form factors in flavor-SU(3) appears promising.

\label{sec:Summary}

\section{Acknowledgments}
We thank Gunnar Bali, John Bulava, Daniel Mohler, and Thomas Wurm for helpful discussions. M.F.M. Lutz thanks Denis Bertini for support on distributed
computing issues

\newpage
\section{Appendix A}
In the course of evaluating the axial-vector form factor of the nucleon 
the following scalar one-loop integrals occur:
\begin{eqnarray}
%%%%%%%%%%%%%%%%%%%%
&& A_{f,i}^{1,k}=\int \frac{d^d l}{(2\pi)^{d}}\,
\frac{\mu^{4-d}(l\cdot \bar{p})^f\left(l^2\right)^k(l\cdot p)^i}{(l-p)^2-M_R^2}\,,
\nonumber\\
%%%%%%%%%%%%%%%%%%%%
&& A_{f,i}^{2,k}=\int \frac{d^d l}{(2\pi)^{d}}\,
\frac{\mu^{4-d}(l\cdot \bar{p})^f\left(l^2\right)^k(l\cdot p)^i}{l^2-m_Q^2}\,,
\nonumber\\
%%%%%%%%%%%%%%%%%%
&& A_{f,i}^{3,k}=\int \frac{d^d l}{(2\pi)^{d}}\,
\frac{\mu^{4-d}(l\cdot \bar{p})^f\left(l^2\right)^k(l\cdot p)^i}{(l-\bar{p})^2-M_L^2}\,,
\nonumber\\
%%%%%%%%%%%%%%%%%%%%%%%
&& B_{f,i}^{1,k}=\int \frac{d^d l}{(2\pi)^{d}}\,
\frac{\mu^{4-d}(l\cdot \bar{p})^f\left(l^2\right)^k(l\cdot p)^i}{((l-\bar{p})^2-M_L^2)(l^2-m_Q^2)}\,,
\nonumber\\
%%%%%%%%%%%%%%%%%%%%%
&& B_{f,i}^{2,k}=\int \frac{d^d l}{(2\pi)^{d}}\,
\frac{\mu^{4-d}(l\cdot \bar{p})^f\left(l^2\right)^k(l\cdot p)^i}{((l-\bar{p})^2-M_L^2)
((l-p)^2-M_R^2)}\,,
\nonumber\\
%%%%%%%%%%%%%%%%%%%%%
&& B_{f,i}^{3,k}=\int \frac{d^d l}{(2\pi)^{d}}\,
\frac{\mu^{4-d}(l\cdot \bar{p})^f\left(l^2\right)^k(l\cdot p)^i}{(l^2-m_Q^2)
((l-p)^2-M_R^2)}\,,
\nonumber\\
%%%%%%%%%%%%%%%%%%%%
&& C_{f,i}^{k}=\int \frac{d^d l}{(2\pi)^{d}}
\frac{\mu^{4-d}(l\cdot \bar{p})^f\left(l^2\right)^k(l\cdot p)^i}{((l-\bar{p})^2-M_L^2)(l^2-m_Q^2)
((l-p)^2-M_R^2)}\,,
\label{triangledef} 
\end{eqnarray}
with the space-time dimension $d$ and the renormalization scale $\mu$ of dimensional regularization. 
In this Appendix we present a convenient recursion scheme in terms of which all such integrals 
can be systematically expressed in the Passarino-Veltman basis,
\begin{eqnarray}
&& I_R=\int \frac{d^dl}{(2\pi)^d}\,\frac{i\,\mu^{4-d}}{l^2-M_R^2}=i\,A_{0,0}^{1,0} \,,
\hspace{2cm}
I_Q=\int \frac{d^dl}{(2\pi)^d}\,\frac{i\,\mu^{4-d}}{l^2-m_Q^2}=i\,A_{0,0}^{2,0} \,,
\nonumber\\
&&I_{L}=\int \frac{d^dl}{(2\pi)^d}\,\frac{i\mu^{4-d}}{l^2-M_{L}^2}=i\,A_{0,0}^{3,0} \,,
\hspace{0.4cm}
\nonumber\\
&&I_{LQ}\left(\bar{p}^2\right)=\int \frac{d^dl}{(2\pi)^d}\,
\frac{-i\,\mu^{4-d}}{((l-\bar{p})^2-M_{L}^2)(l^2-m_Q^2)}=-i\,B_{0,0}^{1,0} \,,
\nonumber\\
&&I_{LR}\left(t\right)=\int \frac{d^dl}{(2\pi)^d}\,
\frac{-i\,\mu^{4-d}}{((l-(\bar{p}-p))^2-M_{L}^2)(l^2-M_R^2)}=-i\,B_{0,0}^{2,0}\,,
\nonumber\\
&&I_{QR}\left(p^2\right)=\int \frac{d^dl}{(2\pi)^d}\,
\frac{-i\mu^{4-d}}{(l^2-m_Q^2)((l-p)^2-M_R^2)}=-i\,B_{0,0}^{3,0}\,,
\nonumber\\
&&I_{LQR}\left(\bar{p}^2,p^2\right)=\int \frac{d^dl}{(2\pi)^d}\,
\frac{i\,\mu^{4-d}}{((l-\bar{p})^2-M_{L}^2)(l^2-m_Q^2)((l-p)^2-M_R^2)}=i\,C_{0,0}^{0}\,.
\end{eqnarray}
Our scheme goes in three steps. First we consider the class of triangle integrals $C_{f,i}^k$. Given the 
relations
\begin{eqnarray}
&& C_{f,i}^k=\frac{v_R^2}{2}\,C_{f,i-1}^k-\frac12 B_{f,i-1}^{1,k} +\frac12\, B_{f,i-1}^{2,k}\,, 
\qquad \qquad C_{f,i}^k=m_Q^2\,C_{f,i}^{k-1}+  B_{f,i}^{2,k-1}\,,
\nonumber\\
&& C_{f,i}^k=\frac{v_L^2}{2}\,C_{f-1,i}^k-\frac12 \,B_{f-1,i}^{3,k} +\frac12 \,B_{f-1,i}^{2,k}\,,
\label{C-rules}
\end{eqnarray}
with
\begin{align}
v_R^2=p^2-M_R^2+m_Q^2, \hspace{1.5cm} v_L^2=\bar{p}^2-M_L^2+m_Q^2\,,
\hspace{1.5cm} v_C^2=q^2-M_R^2+M_L^2 \,,
\label{vDefinition}
\end{align}
for any values of $f,i$ and $k$ the  function $C_{f,i}^k$ can be expressed in terms of 
$C_{0,0}^0$ and the set of scalar bubble functions $B_{f,i}^{1-3,k}$. 

In the second step we use three sets of recurrence relations. The first two read 
\begin{eqnarray}
&& B_{f,i}^{1,k}=\frac{v^2_L}{2}B^{1,k}_{f-1,i}-
\frac12\, A^{2,k}_{f-1,i}+\frac12\, A^{3,k}_{f-1,i}\,, \qquad \qquad
B^{1,k}_{f,i}=m_Q^2\,B^{1,k-1}_{f,i}+A^{3,k-1}_{f,i},
\nonumber\\
&& B_{0,i}^{1,0}=
\sum_{k=0/1}^i {i\choose k}(i-k-1)!!\,(p\cdot \bar{p})^k \,(p^2)^{(i-k)/2}\,
\Big(a_{L,k}^{(i)}\,A^{2,0}_{0,0}+b_{L,k}^{(i)}\,A^{3,0}_{0,0}+
c_{L,k}^{(i)}\,B^{1,0}_{0,0}\Big)\,,
\nonumber\\
%%%%%%%%%%%%%%%%
&& B_{f,i}^{3,k}=\frac{v^2_R}{2}\,B^{3,k}_{f,i-1}-
\frac12 \,A^{2,k}_{f,i-1}+\frac12\, A^{1,k}_{f,i-1}\,,\qquad \qquad B^{3,k}_{f,i}=m_Q^2\,B^{3,k-1}_{f,i}+A^{1,k-1}_{f,i},
\nonumber\\
&&B_{f,0}^{3,0}=
\sum_{k=0/1}^f {f\choose k}(f-k-1)!!\,(\bar{p}\cdot p)^k\,(\bar{p}^2 )^{(f-k)/2}\,
\Big(a_{R,k}^{(f)}\,A^{2,0}_{0,0}+b_{R,k}^{(f)}\,A^{1,0}_{0,0}+
c_{R,k}^{(f)}\,B^{3,0}_{0,0}\Big)\,,
\label{bubble13down}
\end{eqnarray}
where the factors $a_{R,k}^{(f)}$, $b_{R,k}^{(f)}$ and $c_{R,k}^{(f)}$  are derived  in Appendix B. Note that the sum over $k$ in \eqref{bubble13down} runs in steps of 2.
The coefficients $a_{L,k}^{(f)}$, $b_{L,k}^{(f)}$ and $c_{L,k}^{(f)}$
are obtained by the replacement $R\to L$ in the coefficients
$a_{R,k}^{(f)}$, $b_{R,k}^{(f)}$ and $c_{R,k}^{(f)}$. 
It remains to provide a recursion relation for the $B_{f,i}^{2,k}$ bubbles. We find
\begin{eqnarray}
&& B_{f,i}^{2,k}=\frac{1}{2}\,B^{2,k+1}_{f-1,i}+\frac{\bar{p}^2-M_L^2}{2}\,B^{2,k}_{f-1,i}-
\frac12 \,A^{1,k}_{f-1,i}\,, \qquad 
\nonumber\\
&& B_{f,i}^{2,k}=\frac{1}{2}\,B^{2,k+1}_{f,i-1}+\frac{p^2-M_R^2}{2}\,B^{2,k}_{f,i-1}-
\frac12 \,A^{3,k}_{f,i-1}\,,
\nonumber\\
&& B_{0,0}^{2,k}= \sum_{v,w,z=0}^\infty\,K^{(k)}_{vwz}\,A^{1,v}_{w+z,0}
+ \sum_{n=0}^k  \sum_{y=0/1}^n \,K^{(k)}_{ny}\,
\Big(a_{C,y}^{(n)}\,A^{3,0}_{0,0}+
b_{C,y}^{(n)}\,A^{1,0}_{0,0}+c_{C,y}^{(n)}\,B^{2,0}_{0,0}\Big) \,,
\nonumber\\  \nonumber\\
&& K^{(k)}_{vwz} =
\sum_{n,j=0}^\infty {k\choose n}\,{k-n\choose j}\,
\sum_{u=0}^{n-1}\,{u\choose v}\,{u-v\choose w}\, {j\choose z}
\nonumber\\
&& \qquad \;\; \times \,(-1)^{j-z+w}
(\bar{p}^2)^{k-n-z+u-v-w} \,(M_L^2)^{n-1-u}\,2^{j+w}\,,
\nonumber\\
&& K^{(k)}_{ny} = 2^n\,(\bar{p}^2+M_L^2)^{k-n} \,{n\choose y}(n-y-1)!!\,(\bar{p}\cdot p-\bar{p}^2)^y\,(\bar{p}^2)^{(n-y)/2} \,,
\label{bubble2down}
\end{eqnarray}
where we use the convention ${a\choose b}=0$ for $b > a$. The coefficients $a_{C,k}^{(f)}$, $b_{C,k}^{(f)}$ and $c_{C,k}^{(f)}$
follow from $a_{R,k}^{(f)}$, $b_{R,k}^{(f)}$ and $c_{R,k}^{(f)}$
by the replacements $p\to q$ and $m_Q^2\to M_L^2$.
By means of the recurrence relations (\ref{bubble13down}) and (\ref{bubble2down})
any $B_{f,i}^{1-3,k}$ can be expressed in terms of $B_{0,0}^{1-3,0}$ and the set of tadpole integrals $A_{f,i}^{1-3,k}$. 

In our final step we need to apply the following recurrence relations for the tadpole integrals. We find
\begin{eqnarray}
&& A_{f,i}^{1,k}= \sum_{j,w,h=0}^\infty \,K^{(f,k,i)}_{jwg}
 h_{R,0}^{(j+w+g)}\,A^{1,0}_{0,0}\,, \qquad \qquad 
 A_{f,i}^{3,k}=  \sum_{j,w,h=0}^\infty \bar K^{(f,k,i)}_{jwg}\,h_{L,0}^{(j+g+w)}\,A^{3,0}_{0,0} \,, 
\nonumber\\
&& A_{f,i}^{2,k}=m_Q^2\,A_{f,i}^{2,k-1} \,, \qquad \qquad \qquad \qquad \qquad \quad
A_{f,i}^{2,0}= K_{f,i}\,h_{Q,0}^{(i+f)}\,A^{2,0}_{0,0}\,, \qquad 
\nonumber\\ \nonumber\\
&&K^{(f,k,i)}_{jwg} = {f\choose g} \,{i\choose j} \,
\sum_{v=0}^k {k\choose v} \,{k-v\choose w}\,
2^{w}\,(p^2)^{i-j+v}\,
(p\cdot \bar{p})^{f-g}\,(M_R^2)^{k-v-w}\,
\sum_{u=0/1}^{\rm{Min}(j+w,g)}
{g\choose u} 
\nonumber\\
&& \qquad \qquad \quad\times {j+w\choose u}\,u! \,
(j+w-u-1)!!\, (g-u-1)!!\,
(p^2)^{(j+w-u)/2}\,(\bar{p}^2 )^{(g-u)/2}(p\cdot \bar{p})^u \,,
\nonumber\\
&& \bar K^{f,k,i}_{jwg} = {f\choose g} \,{i\choose j}
\sum_{v=0}^k {k\choose v} \,{k-v\choose w}
2^{w}\left(\bar{p}^2\right)^{f-g+v}
(p\cdot \bar{p})^{i-j}\,\left(M_L^2\right)^{k-v-w}\,
\sum_{u=0/1}^{\rm{Min}(j,g+w)}
{g+w\choose u}
\nonumber\\
&& \qquad \qquad \quad \times {j\choose u}\,u!\,(j-u-1)!!\,(g+w-u-1)!! \,
(p^2)^{(j-u)/2}\,(\bar{p}^2)^{(g+w-u)/2}\,(p\cdot \bar{p})^u \,,
\nonumber\\
&& K_{f,i} =\sum_{u=0/1}^{{\rm Min}(i,f)}\,
{f\choose u}{i\choose u}\,u!\,
(i-u-1)!!\,(f-u-1)!!\,
(p^2)^{(i-u)/2}\,(\bar{p}^2)^{(f-u)/2}\,(p\cdot \bar{p})^u \,,
\nonumber\\
&& h_{Q,k}^{(f)}=\prod_{i=0}^{(f-k-1)/2}\frac{m_Q^{f-k}}{(d+2\,i)}\,,
\end{eqnarray}
which upon iteration leads to explicit results for the tadpole integrals  $A_{f,i}^{1-3,k}$
in terms of $A_{0,0}^{1-3,0}$.

\section{Appendix B}
Our results for $a_{R,k}^{(f)}$, $b_{R,k}^{(f)}$ and $c_{R,k}^{(f)}$ are determined  by
the following recursion relations:
\begin{eqnarray}\label{recrelfactora1}
&& \frac{v_R^4}{4}\,a_{R,k}^{(f)}=
\frac12\,h_{Q,0}^{(f+1)}\,\delta_{k1}+
\frac{v_R^2}{4}\,h_{Q,0}^{(f)}\,\delta_{k0}+
 (2\,k+1)\,p^2\, a_{R,k}^{(f+2)} +
k\,(k-1)\,a_{R,k-2}^{(f+2)}+
p^4\, a_{R,k+2}^{(f+2)}\,,
\nonumber\\
&&  m_Q^2\,a_{R,k}^{(f)}=(k+f+d)\,a_{R,k}^{(f+2)}+ p^2\, a_{R,k+2}^{(f+2)} \,,
\nonumber\\ \nonumber\\
&& \frac{v_R^4}{4}\,b_{R,k}^{(f)}=
-\frac{v_R^2}{4}\,h_{R,k}^{(f)}-
\frac12\,p^2\,h_{R,k+1}^{(f+1)}
-\frac12\,k \,h_{R,k-1}^{(f+1)}
\nonumber\\
&& \qquad \quad \;\,+\, (2\,k+1)\,p^2 \,b_{R,k}^{(f+2)} +
k(k-1)\,b_{R,k-2}^{(f+2)}+
p^4\, b_{R,k+2}^{(f+2)}\,,
\nonumber\\
&& m_Q^2\,b_{R,k}^{(f)}=-h_{R,k}^{(f)}
+ (k+f+d)\,b_{R,k}^{(f+2)}+ p^2\, b_{R,k+2}^{(f+2)}\,,
\nonumber\\ \nonumber\\
&& \frac{v_R^4}{4}\,c_{R,k}^{(f)}=
 (2\,k+1)\,p^2 \,c_{R,k}^{(f+2)} +
k\,(k-1)\,c_{R,k-2}^{(f+2)}+
p^4 \,c_{R,k+2}^{(f+2)}\,,
\nonumber\\
&& m_Q^2\,c_{R,k}^{(f)}=(k+f+d)\,c_{R,k}^{(f+2)}+ p^2\, c_{R,k+2}^{(f+2)}\,,
\label{res-abc-iteration}
\end{eqnarray}
once supplemented by the start values for  $f=0$ and $f=1$ 
\begin{eqnarray}
&& a_{R,0}^{(0)}=0\,,\hspace{1.8cm}b_{R,0}^{(0)}=0\,,\hspace{1.3cm}c_{R,0}^{(0)}=1 \,,\hspace{1cm}
\nonumber\\
&&a_{R,1}^{(1)}=-\frac{1}{2\,p^2}\,,\hspace{1cm}b_{R,1}^{(1)}=\frac{1}{2\,p^2}\hspace{1cm}
c_{R,1}^{(1)}=\frac{v_R^2}{2\,p^2}\,,\hspace{1cm}
\label{res-abc-start}
\end{eqnarray}
with $v_R^2=p^2-M_R^2+m_Q^2$. 
We derived explicit expressions for $f=2,3,4$ as implied by 
\eqref{res-abc-iteration} and \eqref{res-abc-start} and show the non-zero contributions. It holds that
\begin{eqnarray}
&& a_{R,0}^{(2)}=\frac{v_R^2}{4\,p^2\,(d-1)} \,,\qquad \qquad \qquad \qquad \qquad \quad
 a_{R,2}^{(2)}=\frac{-d\,v_R^2}{4\,p^4\,(d-1)} \,,
\nonumber\\
&& b_{R,0}^{(2)}=\frac{-1}{p^2\,(d-1)}\left(\frac{v_R^2}{4}-\frac{p^2}{2}\right) \,,
\qquad \qquad  \quad \quad \;
 b_{R,2}^{(2)}=\frac{1}{p^4\,(d-1)}\left(\frac12\,d\,p^2+\frac14\,d\,v_R^2-p^2
\right) \,,
\nonumber\\
&& c_{R,0}^{(2)}=
\frac{-1}{p^2\,(d-1)}
\left(\frac{v_R^4}{4}-p^2\,m_Q^2\right)
\,,\qquad \quad \quad\; \;\;
c_{R,2}^{(2)}=\frac{1}{p^4\,(d-1)}\left(\frac{1}{4}\,d\,v_R^4-p^2\,m_Q^2\right) \,,
\nonumber\\ \nonumber\\
&&a_{R,1}^{(3)}=\frac{-1}{(d-1)\,p^2}
\left(-\frac{v_R^4}{8\,p^2}-\frac{m_Q^2}{2\,d}+\frac{m_Q^2}{2}\right) \,,\qquad\! \!
a_{R,3}^{(3)}=\frac{-1}{(d-1)\,p^4}
\left((d+2)\frac{v_R^4}{8\,p^2}+\frac{m_Q^2}{d}-m_Q^2\right)
\nonumber\\
&&b_{R,1}^{(3)}=\frac{-1}{(d-1)}
\left(-\frac{1}{2}+\frac{v_R^4}{8\,p^4}+\frac{v_R^2}{4\,p^2}-\frac{m_Q^2}{2\,p^2}+
\frac{M_R^2}{2\,d\,p^2}\right) \,,\qquad
\nonumber\\
&& b_{R,3}^{(3)}=\frac{-1}{p^4\,(d-1)}
\left(p^2\left(2-\frac{d}{2}\right)+\frac{3}{2}\,m_Q^2-\frac{v_R^4}{8\,p^2}(2+d)-
\frac{v_R^2}{2}(2+d)-\frac{M_R^2}{2\,d}(2+d)\right) \,,
\nonumber\\
&&c_{R,1}^{(3)}=\frac{1}{(d-1)\,p^2}
\left(\frac{m_Q^2\,v_R^2}{2}-\frac{v_R^6}{8\,p^2}\right) \,,\qquad \qquad
c_{R,3}^{(3)}=\frac{1}{(d-1)\,p^4}
\left(-\frac{3\,m_Q^2\,v_R^2}{2}+\frac{d\,v_R^6}{8\,p^2}+\frac{v_R^6}{4\,p^2}\right) \,,
\nonumber\\ \nonumber\\
&& a_{R,0}^{(4)}=\frac{-v_R^2}{16\,d\,(d^2-1)\,p^4}
\Big(d\,v_R^4-8\,d\,m_Q^2\,p^2+4\,m_Q^2\,p^2\Big)
\,, \nonumber\\ &&
a_{R,2}^{(4)}=\frac{-v_R^2}{16\,d\,(d^2-1)\,p^6}
\Big(-\,d\,(d+2)\,v_R^4-4\,d\,m_Q^2\,p^2-8\,m_Q^2\,p^2\Big)
\,, \nonumber\\ &&
a_{R,4}^{(4)}=\frac{-v_R^2\,(d+2)}{16\,d\,(d^2-1)\,p^8}
\Big(\,d\,(d+4)\,v_R^4-20\,d\,m_Q^2\,p^2+16\,m_Q^2\,p^2\Big)
\,, \nonumber\\ &&
b_{R,0}^{(4)}=\frac{v_R^2-2\,p^2}{16\,d\,(d^2-1)\,p^4}\,\Big(4\,M_R^2\,p^2+
\,d\,\big((m_Q^2-M_R^2)^2-2\,p^2\,(m_Q^2+3\,M_R^2)+p^4\big)\Big)
\,, \nonumber\\ &&
b_{R,2}^{(4)}=
\Big(p^4 \,\big(d^2 \,(m_Q^2+11\,   M_R^2 )-2\,d \,(m_Q^2+3\,
   M_R^2 )-8 \,M_R^2\big)
\nonumber\\
&& \qquad \!   -\,p^2 \,
   (m_Q^2-M_R^2 )\, \big(d^2 \,
   (m_Q^2-5 M_R^2 )-2\,d\,
   (m_Q^2+3 M_R^2 )+8\,   M_R^2\big)
\nonumber\\
&& \qquad \!-\,d \,(d+2)\,
   (m_Q^2-M_R^2 )^3+d \,(d+2)\,
   p^6\Big)/\Big((16\,d \,\big(d^2-1\big)\, p^6\Big)
\,, \nonumber\\ &&
b_{R,4}^{(4)}=
\Big( (11\, d^2-6\,d-8)\, p^4 \,\big(d\,
   (m_Q^2-M_R^2)-4\,M_R^2\big)
 \nonumber\\
 && \qquad + \, (5\, d^2+6\,d-8 )
   p^2 \,(m_Q^2-M_R^2)\,
   \big(d \,(m_Q^2-M_R^2)-4\, M_R^2\big)
\nonumber\\
 && \qquad +\, d\, (d^2+6\,d+8 )
   (m_Q^2-M_R^2)^3+3 \,d \,(5
   d^2-2\,d-8)\, p^6\Big)/ \big(16\,d \,(d^2-1)\,
   p^8\big)
\,, \nonumber\\ &&
c_{R,0}^{(4)}=\frac{(v_R^4-4\,m_Q^2\,p^2)^2}{16\,(d^2-1)\,p^4}
\,, \qquad \qquad
c_{R,2}^{(4)}=\frac{v_R^4-4\,m_Q^2\,p^2}{-16\,(d^2-1)\,p^6}
\Big((d+2)\,v_R^4-4\,m_Q^2\,p^2\Big)
\,, \nonumber\\ &&
c_{R,4}^{(4)}=\Big( (d^2+6\,d+8)\,
   (m_Q^2-M_R^2 )^4
\nonumber\\ 
&& \qquad +\,6 \,p^4\,
   \big(d^2 \,(m_Q^2-M_R^2 )^2-2\,d\,
   \big(m_Q^4+2\, m_Q^2 \,M_R^2-3\,   M_R^4\big)+8 \,M_R^4\big)
\nonumber\\
&& \qquad   +\,(d^2+6\, d+8 )\, p^8+4 \,(d+2)\, p^6
   \big((d-2)\,m_Q^2-(d+4)\, M_R^2\big)
\nonumber\\
&& \qquad 
   + \,4\,   (d+2) \,p^2\,(m_Q^2-M_R^2)^2\, \big((d-2)\,
   m_Q^2-(d+4) \,M_R^2\big)\Big)/\big(16\, (d^2-1) \,p^8\big)\,.
\end{eqnarray}

\section{Appendix C}
We link the triangle integral and its derivative at $t=0$ to  renormalized bubble integrals, which implies

\begin{eqnarray}\label{ILpiRinbubbles}
&&\bar{I}_{L\pi R}(t=0)=\frac{\bar{I}_{\pi L}(M_N)-\bar{I}_{\pi R}(M_N)}{M_R^2-M_L^2}
+\frac{\log\big[\frac{M_R^2}{M_L^2}\big]-\gamma_{N}^L+\gamma_{N}^R}
{16\pi^2\,(M_R^2-M_L^2)}
-\gamma_{L\pi R}\,,
\nonumber \\ \nonumber
&&\bar{I}_{R\pi R}(t=0)=-\frac{\partial \bar{I}_{\pi R}(M_N)}{\partial M_R^2}+\frac{1}{16\pi^2\,M_R^2}-\gamma_{R\pi R}
\nonumber\\
&&\qquad \qquad
=\frac{2\, \bar{I}_{\pi|R}  + (M_N^2+m_\pi^2-M_R^2)\,(\bar{I}_{\pi R}-\frac{\gamma_N^R}{16\pi^2})}{((M_N-M_R)^2-m_\pi^2)\,((M_N+M_R)^2-m_\pi^2)}-\gamma_{R\pi R}\,,
\end{eqnarray}
and
\begin{eqnarray}
 &&\frac{d \bar{I}_{L\pi R}(M_N^2,M_N^2,t=0)}{dt}=
 \frac{1}{96\pi^2 M_N^2 (M_L^2-M_R^2)^3}
\Big\{\frac{1}{2}\, \big(M_L^4 + 4\, M_N^4 + 4 \,m_\pi^4 + M_R^4 - 8 \,M_N^2m_\pi^2 
\nonumber\\ 
&& \quad -4\, M_N^2M_R^2 
  -4\, M_L^2M_N^2  + 2\, M_L^2M_R^2 \big)\,
  \log\left[\frac{M_R^2}{M_L^2}\right]
  - 4\,m_\pi^2\, \Big(M_L^2\log\left[\frac{M_R^2}{m_\pi^2}\right]-M_R^2\log\left[\frac{M_L^2}{m_\pi^2}\right]\Big)
  \Big\}
 \nonumber\\
 &&+\,\frac{1}{6 M_N^2 (M_L^2-M_R^2)^3}
 \Big\{ \big(M_R^4 -M_L^4-2 M_L^2 M_N^2+2 M_L^2 m_\pi^2+2\, M_N^2 M_R^2-2\, m_\pi^2 M_R^2 \big)(\bar{I}_{LR}-\gamma_{LR})
  \nonumber\\ 
 && \quad  +\,\big(-M_L^4-M_L^2 M_N^2-M_L^2 m_\pi^2+3\, M_L^2 M_R^2+2 M_N^4-4\, M_N^2 m_\pi^2-3\,M_N^2 M_R^2
 \nonumber\\ 
 && \quad +\,2\, m_\pi^4-3 \,m_\pi^2 M_R^2 \big)\,\big(\bar{I}_{\pi L}-\gamma^N_L \big)
  +\big(3 \,M_L^2 M_N^2+3\, M_L^2 m_\pi^2-3\, M_L^2 M_R^2-2 \,M_N^4+4\, M_N^2 m_\pi^2
 \nonumber\\ 
 && \quad +\,M_N^2 M_R^2-2\, m_\pi^4+m_\pi^2 M_R^2+M_R^4\big)\,(\big(\bar{I}_{\pi R}-\gamma^N_R \big)\Big\}\,.
\label{dILpiRdtinbubbles}
\end{eqnarray}

The Feynman parameterization of triangle integral \eqref{res-gammaLpiR} reads
\begin{eqnarray}\label{ILpiRFeynman}
&&\bar{I}_{L\pi R}(t)=-\int_0^1\int_0^{1-u}  dv\, du \, \frac{1}{(4\pi)^2\Omega^2(t)}-\gamma_{L\pi R}\,,
\\
&& \Omega^2(t)=-m_\pi^2+v((1-v)M_N^2-M_R^2+m_\pi^2)+u((1-u)M_N^2-M_L^2+m_\pi^2)+uv(t-2M_N^2)\,.\nonumber
\end{eqnarray}

\section{Appendix D}
\allowdisplaybreaks[1] 
We table the factors $\alpha_{ab}$ introduced in Eq. \eqref{K-one-loop}:

\begin{eqnarray}
  &&\alpha_{01}=\frac{12\,M^4+26\,M^3\,\Delta+18\,M^2\,\Delta^2+6\,M\,\Delta^3+\Delta^4}{12\,M^2\,(M+\Delta)^2}
\,,  \nonumber\\
  &&\alpha_{02}=\frac{(2\,M+\Delta)^2\,(6\,M^4+13\,M^3\,\Delta+18\,M^2\,\Delta^2+12\,M\,\Delta^3+2\,\Delta^4)}{24\,M^2\,(M+\Delta)^4}
\,,  \nonumber\\
  &&\alpha_{03}=\frac{(2\,M+\Delta)^3\,(5\,M+\Delta)}{40\,M^2\,(M+\Delta)^2}
\,,  \nonumber\\
  &&\alpha_{10}=\alpha_{11}=\frac{(2\,M+\Delta)^2\,(5\,M^2+5\,M\,\Delta+\Delta^2)}{20\,M^2\,(M+\Delta)^2}
\,,  \nonumber\\
  &&\alpha_{12}=\frac{20\,M^3-66\,M^2\,\Delta-66\,M\,\Delta^2-9\,\Delta^3}{20\,M\,(M+\Delta)^2}
\,,  \nonumber\\
  &&\alpha_{13}=\frac{20\,M^5+60\,M^4\,\Delta+87\,M^3\,\Delta^2+65\,M^2\,\Delta^3+23\,M\,\Delta^4+3\,\Delta^5}{20\,M^2\,(M+\Delta)^3}
   \,,  \nonumber\\
    &&\alpha_{20}=\frac{(2\,M+\Delta)^2\,(15\,M^3+31\,M^2\,\Delta+19\,M\,\Delta^2+4\,\Delta^3)}{60\,M^3\,(M+\Delta)^2}
 \,,  \nonumber\\
    &&\alpha_{21}=\frac{(2\,M+\Delta)^2\,(5\,M^2+5\,M\,\Delta+\Delta^2)}{20\,M^2\,(M+\Delta)^2}
\,,  \nonumber\\
  &&\alpha_{22}=\frac{76\,M^4+170\,M^3\,\Delta+170\,M^2\,\Delta^2+73\,M\,\Delta^3+12\,\Delta^4}{76\,M^2\,(M+\Delta)^2}
\,,  \nonumber\\
  &&\alpha_{23}=\frac{60\,M^6+308\,M^5\,\Delta+645\,M^4\,\Delta^2+707\,M^3\,\Delta^3+421\,M^2\,\Delta^4+129\,M\,\Delta^5+16\,\Delta^6}{60\,M^3\,(M+\Delta)^3}
 \,,  \nonumber\\
 &&\alpha_{30}=\frac{(2\,M+\Delta)^3\,(42\,M^4+106\,M^3\,\Delta+129\,M^2\,\Delta^2+72\,M\,\Delta^3+11\,\Delta^4)}{336\,M^3\,(M+\Delta)^4}
\,,  \nonumber\\
 &&\alpha_{31}=\frac{(2\,M+\Delta)^3\,(6\,M^4+114\,M^3\,\Delta+163\,M^2\,\Delta^2+48\,M\,\Delta^3+5\,\Delta^4)}{48\,M^3\,(M+\Delta)^4}
\,,  \nonumber\\
  &&\alpha_{32}=\frac{(2\,M+\Delta)\,(4\,M^5+268\,M^4\,\Delta+538\,M^3\,\Delta^2+449\,M^2\,\Delta^3+188\,M\,\Delta^4+29\,\Delta^5)}{8\,M^2\,(M+\Delta)^4}
\,,  \nonumber\\
  &&\alpha_{33}=\frac{(42\,M^6+170\,M^5\Delta+440\,M^4\Delta^2+599\,M^3\Delta^3+436\,M^2\Delta^4+163\,M\,\Delta^5+22\,\Delta^6)}{168\,M^3\,(M+\Delta)^5/(2\,M+\Delta)^2}
   \,,  \nonumber\\
        &&\alpha_{40}=\frac{(2\,M+\Delta)^4\,(5\,M+\Delta)}{80\,M^3\,(M+\Delta)^2}
\,,  \nonumber\\
   &&\alpha_{41}=0
\,,  \nonumber\\
  &&\alpha_{42}=\frac{(2\,M+\Delta)^2\,(20\,M^3+24\,M^2\,\Delta+19\,M\,\Delta^2+3\,\Delta^3)}{80\,M^3\,(M+\Delta)^2}
\,,  \nonumber\\
  &&\alpha_{43}=\frac{(2\,M+\Delta)^3\,(20\,M^3+36\,M^2\,\Delta+29\,M\,\Delta^2+5\,\Delta^3)}{160\,M^3\,(M+\Delta)^3} 
  \,,  \nonumber\\
  &&\alpha_{50}=\alpha_{51}=\frac{(2\,M+\Delta)^3\,(5\,M^2+5\,M\,\Delta+\Delta^2)}{40\,M^3\,(M+\Delta)^2}
 \,,  \nonumber\\
  &&\alpha_{52}=\frac{(2\,M+\Delta)\,(48\,M^4+88\,M^3\,\Delta+76\,M^2\,\Delta^2+28\,M\,\Delta^3+5\,\Delta^4)}{96\,M^3\,(M+\Delta)^2}
\,,  \nonumber\\
  &&\alpha_{53}=\frac{(2\,M+\Delta)^2\,(20\,M^4+55\,M^3\,\Delta+63\,M^2\,\Delta^2+31\,M\,\Delta^3+5\,\Delta^4)}{80\,M^3\,(M+\Delta)^3}\,,
  \nonumber\\
     &&\alpha_{60}=\frac{(2\,M+\Delta)^3\,(M^2+M\,\Delta+\Delta)^2}{8\,M^3\,(M+\Delta)^2}
\,,  \nonumber\\
   &&\alpha_{61}=0
\,,  \nonumber\\
  &&\alpha_{62}=\frac{(2\,M+\Delta)^3}{8\,M\,(M+\Delta)^2}
\,,  \nonumber\\
  &&\alpha_{63}=\frac{(2\,M+\Delta)^2\,(4\,M^4+11\,M^3\,\Delta+19\,M^2\Delta^2+15\,M\,\Delta^3+5\,\Delta^4)}{16\,M^3\,(M+\Delta)^3} 
 \,,  \nonumber\\
  &&\alpha_{70}=\frac{(2\,M+\Delta)^4\,(6\,M^2+6\,M\,\Delta+\Delta^2)}{96\,M^4\,(M+\Delta)^2}
\,,  \nonumber\\
 &&\alpha_{71}=\frac{(2\,M+\Delta)^4\,(14\,M^4+10\,M^3\,\Delta+7\,M^2\,\Delta^2+8\,M\,\Delta^3+\Delta^4)}{224\,M^4\,(M+\Delta)^4}
\,,  \nonumber\\
  &&\alpha_{72}=\frac{(2\,M+\Delta)^2\,(12\,M^4+24\,M^3\,\Delta+25\,M^2\,\Delta^2+13\,M\,\Delta^3+2\,\Delta^4)}{48\,M^4\,(M+\Delta)^2}
\,,  \nonumber\\
  &&\alpha_{73}=\frac{(2\,M+\Delta)^3\,(12\,M^4+36\,M^3\,\Delta+43\,M^2\,\Delta^2+21\,M\,\Delta^3+3\,\Delta^4)}{96\,M^4\,(M+\Delta)^3}
  \,.  
  \label{alphas-explicit}
\end{eqnarray}

\newpage
\bibliography{literature} 

\end{document}